\documentclass[epsfig,12pt]{article}
\usepackage{amssymb}
\usepackage{amsfonts}
\usepackage{graphicx}
\usepackage{amsmath}

\input epsf.sty
\newtheorem{theorem}{Theorem}
\newtheorem{acknowledgement}[theorem]{Acknowledgement}

\input{tcilatex}
\makeatletter \@addtoreset{equation}{section}

\begin{document}

\title{\rightline{\mbox{\small
{Lab/UFR-HEP0801/GNPHE/0801}}} \textbf{Generalized MacMahon G}$_{d}\left(
q\right) $\textbf{\ as }\\
\emph{q-deformed} \textbf{CFT}$_{2}$\textbf{\ Correlation Function}}
\author{Lalla Btissam Drissi$^{1,2}${\small \thanks{%
drissilb@gmail.com}}, Houda Jehjouh$^{1,2}${\small \thanks{%
jehjouh@gmail.com}}, El Hassan Saidi$^{1,2,3}${\small \thanks{%
h-saidi@fsr.ac.ma}} \\
%EndAName
{\small 1. Lab/UFR- Physique des Hautes Energies, Facult\'{e} des Sciences,
Rabat, Morocco,}\\
{\small 2. GNPHE, Groupement National de Physique des Hautes Energies, Si%
\`{e}ge focal: FS, Rabat.}\\
{\small 3. Acad\'{e}mie Hassan II des Sciences et Techniques, Coll\`{e}ge
des Sciences Physiques et Chimiques, }\\
{\small Rabat, Morocco.}}
\maketitle

\begin{abstract}
Using $\Gamma _{\pm }\left( z\right) $ vertex operators of the $c=1$ two
dimensional conformal field theory, we give a 2d-quantum field theoretical
derivation of the conjectured d- dimensional MacMahon function G$_{d}\left(
q\right) $. We interpret this function G$_{d}\left( q\right) $ as a $\left(
d+1\right) $- point correlation function $\mathcal{G}_{d+1}\left(
z_{0},...,z_{d}\right) $ of some local vertex operators $\mathcal{O}%
_{j}\left( z_{j}\right) $. We determine these operators and show that they
are particular composites of q-deformed hierarchical vertex operators $%
\Gamma _{\pm }^{\left( p\right) }$, with a positive integer p. In agreement
with literature's results, we find that G$_{d}\left( q\right) $, $d\geq 4$,
cannot be the generating functional of all \textit{d- dimensional }%
generalized Young diagrams .\newline
\end{abstract}

\tableofcontents

\section{Introduction}

\qquad The study of two dimensional (2d) MacMahon function G$_{2}$, and its 
\emph{3d- generalization }G$_{3}$ appear in many areas of statistical
physics, such as crystals growth, crystals melting, Bose-Einstein statistics
and dimer model\textrm{\ \cite{a1}-\cite{a0}}. Recently, these functions
have known a revival of interest in connection with topological string
theory $\mathrm{\cite{b1,b2,b0}}$; in particular in the study of BPS black
holes, given by branes wrapping collapsed cycles in Calabi-Yau orbifolds,
and in the infinite $n$ limit of quiver gauge theories $\mathrm{\cite%
{c1,c3,c4,c5,c0}}$. \textrm{\ }MacMahon functions G$_{2}\left( q\right) $
and G$_{3}\left( q\right) $ are also used in\textrm{\ }the explicit
computation of the amplitudes of A-model topological string on local
Calabi-Yau manifolds \textrm{\cite{1,2,3,4}.}

In \textrm{\cite{3}, }it has been shown that the topological A-model
partition function Z$_{3d}$ on the complex space $\mathbb{C}^{3}$, which
coincides exactly with the 3d- crystal melting partition function Z$%
_{crystal}$, is given by \emph{3d- generalized} MacMahon function G$_{3}$.
This is an important result since topological amplitudes for the full class
of toric Calabi-Yau threefolds X$_{3}$ with \emph{a planar toric geometry}
are recovered just by gluing the $\mathbb{C}^{3}$-vertices $\mathrm{\cite{1}}
$. Amplitudes involving open strings are also recovered up on inserting
special Lagrangian D- branes captured by boundary conditions on the edges of
the 3- vertices \textrm{\cite{4}}. An evidence for a "topological 4-
vertex", in the case of toric Calabi-Yau threefold with \emph{non planar
toric geometry}, has been also studied in \textrm{\cite{5}} and would,
roughly, be described by a \emph{4d- extension} of the generalized MacMahon
function G$_{3}$.

MacMahon functions G$_{2}$ and G$_{3}$ appear as well in representation
theory of infinite dimensional Lie algebras and in topologically twisted $%
U\left( 1\right) $ gauge theories $\mathrm{\cite{g1,g2,g3,g0}}$. They are
respectively the (specialized) character $ch_{R}$ of the basic
representations of the $sl\left( \infty \right) $ and the affine algebra $%
\widehat{sl(\infty )}$ at large central charge $c$ \textrm{\cite%
{g4,g41,g3,g43}}. In the limit $c\rightarrow \infty $, it has been moreover
observed in \textrm{\cite{g5}} that the basic representation of $\widehat{%
sl(\infty )}$ is closely related to the partition function of a three
dimensional free field theory. MacMahon G$_{2}$ and G$_{3}$ are also the
partition functions of the 4d- and 6d-\ topologically twisted U$\left(
1\right) $ gauge theory given by a D4 and D6-branes filling respectively $%
\mathbb{C}^{2}$ and $\mathbb{C}^{3}$ $\mathrm{\cite{c0,c1}}$.

For higher dimensions, it has been checked in \textrm{\cite{7}} that G$%
_{4}\left( q\right) $ cannot be the generating functional of the \emph{4d-
generalized} Young diagrams leading then to the two basic questions: \newline
(\textbf{i}) what is the right generating functional of generalized \emph{d-
dimensional }Young diagrams for $d\geq 4$. \newline
(\textbf{ii}) what is the exact interpretation of G$_{4}\left( q\right) $
and G$_{d}\left( q\right) $ in general.\newline
These questions are not trivial and their answer needs developing more
involved mathematical machinery. Nevertheless, a first step towards the
answer of these questions is to start by deepening the study of the
conjectured generalized \emph{d- dimensional} MacMahon function G$_{d}$. In
particular the issues regarding its interpretation in 2d- quantum field
theory and its explicit derivation using correlation functions of local
vertex operators.

A natural way to reach this goal is to use the "transfer matrix" approach 
\textrm{\cite{3,4,h0} }and borrow ideas from q-deformed QFT$_{2}$\ \cite%
{vafa,mina,Ahl}. This method has been successfully\textrm{\ }used for the
particular case $d=3$ and could, \`{a} priori, extended to higher d-
dimensions. In topological string on local Calabi-Yau threefolds, the key
idea in getting topological closed string partition function relies on
expressing Z$_{3d}^{{\small closed}}$ as a particular vev $\left\langle 0|%
\mathcal{T}|0\right\rangle $ of some hermitian transfer matrix operator $%
\mathcal{T}$. This operator can be factorized as $\mathcal{A}_{+}\mathcal{A}%
_{-}$ with $\mathcal{A}_{+}$ and $\mathcal{A}_{-}$ being \emph{composite}
local vertex operators of the two dimensional $c=1$ conformal field theory.
Implementation of open strings leads to Z$_{3d}^{{\small open}}\sim C_{\nu
\mu \lambda }$ and is achieved as $\left\langle \nu ^{t}|\mathcal{A}%
_{+}\left( \lambda \right) \mathcal{A}_{-}\left( \lambda ^{t}\right) |\mu
\right\rangle $ by inserting boundary states $|\sigma >$ , $\sigma =\nu ,$ $%
\lambda ,$ $\mu $, described by asymptotic 2d- Young diagrams. If we let
string interpretation aside, this construction could be applied as well for
higher d- dimensions where G$_{d}$ is expected to play a central role.

This paper has two main objectives: \newline
\textbf{(1) }Give a conformal field theoretical derivation of the
conjectured \emph{d- dimensional} generalized MacMahon function G$_{d}$
expressed by the following formula,%
\begin{equation*}
G_{d}\left( q\right) =\dprod\limits_{k=1}^{\infty }\left[ \left(
1-q^{k}\right) ^{-\frac{\left( k+d-3\right) !}{\left( k-1\right) !\left(
d-2\right) !}}\right] ,\qquad d\geq 2,
\end{equation*}%
together with the two special ones filling the hierarchy, 
\begin{equation}
G_{1}\left( q\right) =\frac{1}{1-q}\qquad ,\qquad G_{0}\left( q\right) =1.
\label{10}
\end{equation}%
Recall that in combinatorial analysis, the function $G_{3}\left( q\right) $
can be defined as the generating functional of \emph{3-dimensional}
partitions $\Pi ^{\left( 3\right) }$ extending the usual 2d- partitions $\mu
=\Pi ^{\left( 2\right) }$ to higher 3-dimensions. Refined studies regarding G%
$_{4}$ function have revealed that it is not the generating functional of 4d
partitions \textrm{\cite{7,8,4}}. \newline
To fix the ideas, it is interesting to recall that expanding $G_{2}\left(
q\right) $ as a $q^{n}$ power series like, 
\begin{equation*}
G_{2}\left( q\right) =\dprod\limits_{k=1}^{\infty }\left( \frac{1}{1-q^{k}}%
\right) =\sum_{n=0}^{\infty }p_{2}\left( n\right) q^{n},
\end{equation*}%
one gets the number $p_{2}\left( n\right) $ of 2d- partitions (Young
diagrams) containing n boxes. From this view, $G_{2}$\ can be physically
interpreted as the exact partition function $Z_{2}=Tr\left( q^{H}\right) $
of a two dimensional statistical physics system with, 
\begin{equation*}
q=\exp \left( -\frac{1}{KT}\right) ,
\end{equation*}%
and energy spectrum $E_{k}=k$. Here $T$ is the absolute temperature and the
constant $K$ is the Boltzmann one. For instance, $G_{2}\left( q\right) $ is
the partition function of the $c=1$ free Bose gaz. There, the Hamiltonian is
given by $\mathcal{H}=\hbar \omega \sum k\mathcal{N}_{k}$ with $\mathcal{N}%
_{k}=a_{k}^{+}a_{k}$ being the operator number of particles and energy
spectrum $E_{k}=\hbar \omega k$.\newline
Similar expansions can be also made for $G_{d}\left( q\right) $ which then
read as follows%
\begin{equation*}
G_{d}\left( q\right) =\sum_{n=0}^{\infty }p_{d}\left( n\right) q^{n},\qquad
d\geq 3.
\end{equation*}%
For the case $d=3$, the number $p_{3}\left( n\right) $ is precisely the
number of 3d- partitions; but for $d=4$, the number $p_{4}\left( n\right) $
is not the total number of 4d- partitions as it has been explicitly checked
in \textrm{\cite{7}}.\newline
\textbf{(2)} The second objective of the present study is to show that $%
G_{d}\left( q\right) $ can be remarkably interpreted as a $\left( d+1\right) 
$- point correlation function $\mathcal{G}_{d+1}$ of some \emph{q- deformed}
vertex operators $\mathcal{O}_{j}\left( x_{j}\right) $, i.e%
\begin{equation}
G_{d}\left( q\right) =\mathcal{G}_{d+1}\left( x_{0},x_{1},x_{2}\cdots
,x_{d}\right)  \label{re}
\end{equation}%
with $x_{j}=q^{j}$, $j=0,...,d$; \ and 
\begin{equation}
\mathcal{G}_{d+1}=\left\langle 0|\mathcal{O}_{0}\left( x_{0}\right) \mathcal{%
O}_{1}\left( x_{1}\right) \mathcal{O}_{2}\left( x_{2}\right) \cdots \mathcal{%
O}_{d}\left( x_{d}\right) |0\right\rangle .  \label{er}
\end{equation}%
The $\mathcal{O}_{j}\left( x_{j}\right) $'s will be determined in terms of
the usual vertex operators $\Gamma _{\pm }=\exp \left( \Phi _{\pm }\right) $
of the $c=1$ two dimensional bosonic conformal field theory $\mathrm{\cite%
{m1}}$; but also others, denoted like $\Gamma _{\pm }^{\left( p\right) },$
involving q-deformed QFT$_{2}$. This result gives: \newline
(\textbf{i}) a \emph{q-deformed} 2d quantum field theoretical proof of the
conjectured MacMahon function G$_{d}$,\newline
(\textbf{ii}) an interpretation of G$_{d}$ using \emph{q- deformed} $c=1$
conformal field theory rather than CFT$_{2}$ free field theory with central
charge $c\rightarrow \infty $.\newline
(\textbf{iii}) For $d\geq 4$ G$_{d}$ cannot be the generating function of
d-generalized partitions; but rather of a subclass of d- partitions with
very specific boundary conditions.

The organization of this paper is as follows: \newline
In section 2, we introduce the usual vertex operators $\Gamma _{\pm }$ of
the $c=1$\ 2d conformal model and give some of their properties essential
for the next steps. In section 3, we revisit the CFT$_{2}$ derivation of 3d-
generalized MacMahon function G$_{3}$ using transfer matrix method. We also
introduce the q-deformed $\Gamma _{\pm }^{\left( 2\right) }$ vertex
operators. In section 4, we derive the generalized MacMahon function G$_{n}$
for 4d and 5d using transfer matrix method and q-deformed vertex operators $%
\Gamma _{\pm }^{\left( 3\right) }$ and $\Gamma _{\pm }^{\left( 4\right) }$.
In section 5, we give the result for generic \emph{d- dimensions}. In
section 6, we derive $\mathcal{O}_{j}\left( x_{j}\right) $ vertex operators
involved in $G_{d}\left( q\right) $\ re-interpreted as $\left( d+1\right) $-
point correlation function $\mathcal{G}_{d+1}\left( z_{0},...,z_{d}\right) $
in q-deformed $c=1$ CFT$_{2}$. In the conclusion section, we summarize the
main results of the paper accompanied with a discussion. In appendices A and
B, we give more details on the proofs of identities used in the present
study.

\section{Vertex operators: useful properties}

\qquad In this section, we explore some basic properties of the vertex
operators $\Gamma _{\pm }\left( z\right) $ in $c=1$ 2d- conformal field
theory. We study their commutation relations algebra in connection with the
counting of the Hilbert space states and the 2d- partitions (Young
diagrams). We also give special features of $\Gamma _{\pm }\left( z\right) $
which has motivated us to look for the relations (\ref{re}-\ref{er}).

\subsection{Vertex operators in $c=1$ CFT$_{2}$}

\qquad As the\textrm{\ }$c=1$ field vertex operators $\Gamma _{\pm }\left(
z\right) ,$ $z\in C,$ have been well studied and are quite known in 2d
conformal field theory \textrm{\cite{06,07,h0}}, we shall come directly to
the main points by considering the three following materials needed for the
study of $\Gamma _{\pm }\left( z\right) $ and their extensions to be
considered in this study: \newline
\textbf{(1)} $U\left( 1\right) $ \textbf{Kac-Moody algebra}. \newline
In CFT$_{2}$ on the complex line $C$ parameterized by the coordinate $z$,
the $U\left( 1\right) $ Kac-Moody algebra is generated by the holomorphic
current $J\left( z\right) $ obeying the following operator product expansion
(OPE)%
\begin{equation}
J\left( z_{1}\right) J\left( z_{2}\right) =\frac{1}{\left(
z_{1}-z_{2}\right) ^{2}}\text{ }+\text{ regular terms.}  \label{a1}
\end{equation}%
Using the Laurent expansion, 
\begin{equation}
J\left( z\right) =\sum_{n\in Z}z^{-n-1}J_{n},\qquad J_{n}=\doint \frac{dz}{%
2i\pi }z^{n}J\left( z\right) ,  \label{a3}
\end{equation}%
the above OPE algebra reads as follows 
\begin{equation}
\left[ J_{n},J_{m}\right] =n\delta _{n+m,0}.  \label{a2}
\end{equation}%
We have, amongst others, $\left( J_{n}\right) ^{\dagger }=J_{-n}$ and $%
J_{n}\left\vert 0\right\rangle =0$ for n$\geq 1$.\newline
\textbf{(2)} c=1 \textbf{conformal model}.\newline
In the 2D conformal field theoretic realization of eqs(\ref{a1}-\ref{a2}),
one distinguishes two free field theoretic realizations of the $c=1$
conformal representation: \newline
\textbf{(i) }The free bosonic realization using a single real (chiral) boson 
$\Phi \left( z\right) $ with the usual two- point correlation function,%
\begin{equation}
\Phi \left( z_{1}\right) \Phi \left( z_{2}\right) =-\ln \left(
z_{1}-z_{2}\right) +\text{ regular terms.}  \label{ln}
\end{equation}%
\textbf{(ii)} The free fermionic realization using a complex one component
fermion $\psi \left( z\right) $. In this case,\textrm{\ }the two- point
correlation function that have a singular term is,%
\begin{equation}
\psi ^{\ast }\left( z_{1}\right) \psi \left( z_{2}\right) =\frac{1}{%
z_{1}-z_{2}}+\text{ regular terms.}  \label{1z}
\end{equation}%
The two- point functions $\psi \left( z_{1}\right) \psi \left( z_{2}\right) $
and $\psi ^{\ast }\left( z_{1}\right) \psi ^{\ast }\left( z_{2}\right) $ are
regular.\newline
The U$\left( 1\right) $ Kac-Moody current $J\left( z\right) $ is given, in
the bosonic representation, by: 
\begin{equation}
J\left( z\right) =\frac{\partial \Phi \left( z\right) }{i\partial z},
\label{a4}
\end{equation}%
while\textrm{\ }it has the following form $J\left( z\right) =:i\psi ^{\ast
}\left( z\right) \psi \left( z\right) :$ in terms of fermions\textrm{.}
Below, we shall mainly focus on the bosonic case; the link with fermionic
representation can be done by using bosonization ideas. \newline
Expanding the 2d chiral scalar field as 
\begin{equation}
\Phi \left( z\right) =\sum_{n\in Z}z^{-n}\Phi _{n}  \label{a5}
\end{equation}%
and rearranging it as $\Phi \left( z\right) =\Phi _{-}\left( z\right) +\Phi
_{0}+\Phi _{+}\left( z\right) ,$ we can write the above expansion as, 
\begin{equation}
\Phi _{-}\left( z\right) =i\sum_{n>0}\frac{1}{n}z^{n}J_{-n},\quad \Phi
_{+}\left( z\right) =-i\sum_{n>0}\frac{1}{n}z^{-n}J_{n}  \label{h1}
\end{equation}%
where we have used 
\begin{equation}
\Phi _{n}=\frac{1}{in}J_{n},\text{\qquad }n\in Z^{\ast }.
\end{equation}%
This identity follows directly by comparing eq(\ref{a4}-\ref{a5}) and (\ref%
{a3}). Notice also that the zero mode $\Phi _{0}$ acts trivially; it will be
ignored in follows.\newline
\textbf{(3)} \textbf{Vertex operators: }$Level$\textbf{\ }\emph{1}.\newline
There are various local field vertex operators that we will encounter in
this present study. The simplest ones, named as \emph{level 1}, are given by 
\begin{equation}
\Gamma _{\pm }\left( z\right) =\exp \Phi _{\pm }\left( z\right) ,\qquad z\in
C,
\end{equation}%
The other vertex operators $\Gamma _{\pm }^{\left( p\right) }\left( z\right) 
$, to be introduced later on, will be named as \emph{level p} vertex
operators. Substituting $\Phi _{\pm }\left( z\right) $ by their expression (%
\ref{h1}), the \emph{Level\ 1} vertex operators ($\Gamma _{\pm }\left(
z\right) \equiv \Gamma _{\pm }^{\left( 1\right) }\left( z\right) $) read
also as follows%
\begin{equation}
\begin{tabular}{llll}
$\Gamma _{-}\left( z\right) $ & $=$ & $\exp \left( i\sum_{n>0}\frac{1}{n}%
z^{n}J_{-n}\right) $ & , \\ 
$\Gamma _{+}\left( z\right) $ & $=$ & $\exp \left( -i\sum_{n>0}\frac{1}{n}%
z^{-n}J_{n}\right) $ & .%
\end{tabular}
\label{ga}
\end{equation}%
These objects may be interpreted as the generating functionals of monomials
of the $J_{m}$ operators. For instance, we have for the leading terms,%
\begin{equation}
J_{-1}=\frac{\partial \Gamma _{-}\left( 0\right) }{i\partial z}\qquad
,\qquad \left( J_{-1}^{2}-iJ_{-2}\right) =\frac{\partial ^{2}\Gamma
_{-}\left( 0\right) }{\left( i\partial z\right) ^{2}}
\end{equation}%
and similar relations for their adjoints. Notice that since the states of
the Hilbert space of $c=1$ conformal theory representation are given by%
\begin{equation}
\dprod\limits_{i\geq 1}\left( J_{-n_{i}}\right) ^{\lambda _{i}}\left\vert
0\right\rangle \qquad ,\qquad J_{n_{i}}\left\vert 0\right\rangle =0,\qquad
n_{i},\text{\ }\lambda _{i}\text{ }\in N\text{ },  \label{bs}
\end{equation}%
it follows that the state%
\begin{equation}
\Gamma _{-}\left( z\right) \left\vert 0\right\rangle \text{ \ ,}  \label{w}
\end{equation}%
is the generating functional of the basis states (\ref{bs}) of the $c=1$ 
\textrm{CFT}$_{2}$ Hilbert space.

\subsection{Algebra of the $\Gamma _{\pm }$ vertex operators}

\qquad The action of the local operators $\Gamma _{\pm }\left( z\right) $\
on the Hilbert space states of the $c=1$ 2d- conformal field theory exhibits
a set of special properties inherited from the algebra of the $J_{\pm n}$
modes (\ref{a2}). Some of these properties are revisited in what follows:%
\newline
(\textbf{1}) The $\Gamma _{\pm }\left( z\right) $ operators obey the
algebra, 
\begin{eqnarray}
\Gamma _{\pm }(x)\Gamma _{\pm }(y) &=&\Gamma _{\pm }(y)\Gamma _{\pm
}(x),\qquad x,y\in C,  \notag \\
\Gamma _{+}(x)\Gamma _{-}(y) &=&\left( 1-\frac{y}{x}\right) ^{-1}\Gamma
_{-}(y)\Gamma _{+}(x),  \label{96}
\end{eqnarray}%
which can be easily established by using eqs(\ref{ln}-\ref{a2}).\newline
(\textbf{2}) The operator $q^{L_{0}}$ acts on $\Gamma _{+}(1)$ and $\Gamma
_{-}(1)$ as a translation operator as shown below, 
\begin{equation}
q^{L_{0}}\Gamma _{\pm }(1)q^{-L_{0}}=\Gamma _{\pm }(q).  \label{sa}
\end{equation}%
This relation will play a crucial role later on, in particular when using
transfer matrix method.\newline
(\textbf{3}) Using the properties $J_{n}|0>=0$ and $<0|J_{-n}=0$ for $n>0$,
we have moreover: \ \newline
(\textbf{i}) For all positions $z$, the operators $\Gamma _{\pm }(z)$ act on
the vacuum as the identity operator: 
\begin{equation}
\Gamma _{+}(z)|0>=|0>,\qquad <0|\Gamma _{-}(z)=<0|.
\end{equation}%
So we have%
\begin{equation}
<0|\Gamma _{-}(z)|0>=1,\qquad <0|\Gamma _{+}(z)|0>=1,  \label{1}
\end{equation}%
and 
\begin{eqnarray}
&<&0|\Gamma _{-}(z)\Gamma _{+}(w)|0>=1,  \notag \\
&<&0|\Gamma _{\pm }(z)\Gamma _{\pm }(w)|0>=1.  \label{2}
\end{eqnarray}%
Notice in passing that, viewing $\Gamma _{\pm }(z)$ as local operator
fields, eqs(\ref{1}) and (\ref{2}) may be interpreted respectively as 1-
point and 2- point Green functions. Notice moreover that since $\Gamma
_{+}(z)$ and $\Gamma _{-}(w)$ are non commuting operators, we have,%
\begin{equation}
<0|\Gamma _{+}(z)\Gamma _{-}(w)|0>\text{ }\neq \text{ }<0|\Gamma
_{-}(z)\Gamma _{+}(w)|0>.
\end{equation}%
We will develop this issue much more later when we come to the derivation of
eqs(\ref{re}-\ref{er}) by using correlation functions.\newline
(\textbf{ii}) As $\Gamma _{-}(z)$ involves all monomials in $J_{-n_{i}}$, 
\begin{equation}
J_{-\mathbf{n}}^{\lambda }\equiv \Pi _{i\geq 1}\left( J_{-n_{i}}\right)
^{\lambda _{i}},
\end{equation}%
where $\lambda =\left( \lambda _{1},\lambda _{2},\ldots \right) $ is a 2d-
partition, the state $\Gamma _{-}(z)|0>$ is reducible and is given by a sum
over all possible 2d- partitions $\lambda $. In particular we have for $z=1$,%
\begin{equation}
\Gamma _{-}(1)|0>=\sum_{\text{{\small 2d partitions} }\lambda }|\lambda >.
\label{222}
\end{equation}%
A similar relation is also valid for $<0|\Gamma _{+}(1)$. More generally,
this relation extends as $\Gamma _{-}(1)|\mu >$ and involves the Schur
function $S_{\mu }^{\text{{\small Schur}}}\left( q\right) $ \textrm{\cite{06}%
}. With these tools, we are in position to proceed for higher dimensional
generalizations.

\section{The 3d- MacMahon function revisited}

Our main objectives here are:\newline
(\textbf{i}) revisit the derivation of Z$_{3d}$\newline
(\textbf{ii}) use CFT$_{2}$ explicit computations to give arguments which
support the existence of a hierarchy of \emph{level p} vertex operators $%
\Gamma _{\pm }^{\left( p\right) }$.

To reach this goal, we first give some details on 3d- partitions (known also
as plane partitions) and its generating functional Z$_{3d}$. Then we present
the explicit computation of the function Z$_{3d}$ using transfer matrix
method. As mentioned in the introduction, Z$_{3d}$ is precisely the
amplitude of the topological 3- vertex of closed strings on $\mathbb{C}^{3}$%
. There, the q- parameter is given by 
\begin{equation}
q=\exp \left( -g_{s}\right) \text{ ,}
\end{equation}%
with $g_{s}$\ being the topological string coupling constant. Z$_{3d}$ is
also the partition function of corner melting 3d- crystals.

\subsection{Plane partitions and 3d- Hilbert states}

\qquad To begin notice that, from the view of combinatory analysis, the3d-
MacMahon function $G_{3d}$ can be defined by the following partition
function 
\begin{equation}
Z_{3d}=\sum_{3d\text{ partitions }\Pi ^{\left( 3\right) }}q^{\left\vert \Pi
^{\left( 3\right) }\right\vert }\text{ },  \label{3dd}
\end{equation}%
where $\left\vert \Pi ^{\left( 3\right) }\right\vert $ is the number of
boxes of the 3d- generalized Young diagram. This relation may be also
written as%
\begin{equation}
Z_{3d}=\sum_{3d\text{ partitions }\Pi ^{\left( 3\right) }}\left\langle \Pi
^{\left( 3\right) }|q^{\mathcal{H}}|\Pi ^{\left( 3\right) }\right\rangle 
\text{ },
\end{equation}%
with 
\begin{equation}
\mathcal{H}\left\vert \Pi ^{\left( 3\right) }\right\rangle =E\left\vert \Pi
^{\left( 3\right) }\right\rangle \text{ },\qquad E=\left\vert \Pi ^{\left(
3\right) }\right\vert \text{ }.
\end{equation}%
The Hilbert space states $\left\vert \Pi ^{\left( 3\right) }\right\rangle $,
to which we shall refer as "\emph{3d- Hilbert states"}, are the quantum
states associated with $\Pi ^{\left( 3\right) }$. The relation (\ref{3dd})
has a remarkable combinatorial interpretation; it is the generating function
of the $p_{3}\left( n\right) $ number of 3d- partitions $\Pi ^{\left(
3\right) }$ with $n$ boxes. The $p_{3}\left( n\right) $ number can be
determined by expanding $Z_{3d}$ like, 
\begin{equation}
Z_{3d}\left( q\right) =\sum_{n=0}^{\infty }p_{3}\left( n\right) q^{n}\text{ }%
,\qquad p_{3}\left( n\right) =\frac{\partial ^{n}Z_{3d}\left( 0\right) }{%
n!\partial q^{n}}\text{ }.
\end{equation}%
Notice also that 3d- partitions $\Pi ^{\left( 3\right) }$ are 3d-
generalizations of Young diagrams and can be decomposed as a sequence%
\footnote{%
3d- partitions $\Pi ^{\left( 3\right) }$ have integer entries $\left( \Pi
_{a,b}\geq 0\right) $ such that $\Pi _{a,b}\geq \Pi _{a+i,b+j}$ $i,j\geq 0$.
These are 3d generalizations of the usual Young diagrams described by the
2d- partitions $\lambda =\left( \lambda _{1},\lambda _{2},...\right) $ with $%
\lambda _{a}\geq \lambda _{a+1}$. The partitions $\Pi ^{\left( 3\right) }$
have several properties; in particular the diagonal slicing in terms of 2d
partitions $\Pi _{t}^{\left( 2\right) }$ used in the transfer matrix method.
The diagonal slicing of $\Pi ^{\left( 3\right) }=\left( \Pi _{a,b}\right) $
is obtained by setting $b=a+t$ where $t\in Z$ parameterizes the sequence $%
\Pi _{t}^{\left( 2\right) }$. For fixed t, $\Pi _{t}^{\left( 2\right) }$ may
be thought of as $\lambda $ with parts $\lambda _{a}=$ $\Pi _{a,a+t}$.} of
2d- partitions $\Pi _{t}^{\left( 2\right) }$like,%
\begin{equation}
\Pi ^{\left( 3\right) }=\sum_{t\in Z}\Pi _{t}^{\left( 2\right) },
\label{dec}
\end{equation}%
where t parameterizes the slices. For fixed integer t, the 2d-partition $\Pi
_{t}^{\left( 2\right) }=\left( \Pi _{a,a+t}\right) _{a\in \mathbb{N}^{\ast
}} $\ lives on the diagonal plane $b=a+t$\ of the cubic lattice $\mathbb{N}%
^{\ast }\times \mathbb{N}^{\ast }\times \mathbb{N}^{\ast }$\ parameterized
by the positive integers $\left( a,b,c\right) $. The diagonal decomposition (%
\ref{dec}) is useful here in the sense it is used in the transfer matrix
method for computing $Z_{3d}$. There exist an other decomposition of $\Pi
^{\left( 3\right) }$\ namely the so called perpendicular decomposition
relevant for the study of the topological vertex. \newline
Expressing the number $\left\vert \Pi ^{\left( 3\right) }\right\vert $ of
boxes of 3d -partition in terms of 2d ones, namely 
\begin{equation}
\left\vert \Pi ^{\left( 3\right) }\right\vert =\sum_{t}\left\vert \Pi
_{t}^{\left( 2\right) }\right\vert ,
\end{equation}%
we can put eq(\ref{3dd}) in the form%
\begin{equation}
Z_{3d}=\sum_{3d\text{ partitions }\Pi ^{\left( 3\right) }}\left(
\dprod\limits_{t}q^{\left\vert \Pi _{t}^{\left( 2\right) }\right\vert
}\right) .
\end{equation}%
To get "3d- generalized Hilbert states" $|\Pi ^{\left( 3\right) }>$, it is
interesting to first recall 2d- generalized Hilbert space states $|\Pi
^{\left( 2\right) }>\equiv |\lambda >$. In the language of the U$\left(
1\right) $ Kac-Moody algebra representations, the Hilbert space states of
the $c=1$ CFT$_{2}$ have the structure%
\begin{equation}
|\lambda >=|\lambda _{1},..,\lambda _{i}...>,\qquad
\end{equation}%
and are completely characterized by 2d- partitions, 
\begin{equation}
\lambda =\left( \lambda _{1},..,\lambda _{i},...\right) ,\qquad \lambda
_{1}\geq \lambda _{2}\geq \cdots ,\qquad \lambda _{i}\in \mathbb{N}.
\end{equation}%
The generating functional of these states is given by $\Gamma _{-}\left(
1\right) |0>=\sum\limits_{\lambda }|\lambda >$, eq(\ref{222}). Generalized
Hilbert space states $|\Pi ^{\left( 3\right) }>$, associated to 3d-
partitions $\Pi ^{\left( 3\right) }$ may be built out 2d - partitions with
interlacing relations \textrm{\cite{3}}. \newline
The generating functional of 3d partitions requires, in the framework of
transfer method, the following \footnote{%
Note that $\Psi _{\pm }\left( 1\right) $\ corresponds to $\Psi _{\pm }\left(
z\right) $ with $z=1$. Note also that $\Psi _{\pm }\left( 1\right) $ depend
on the q- parameter; it has been dropped out for simplicity of notations.} 
\begin{equation}
\Psi _{-}\left( 1\right) |0>=\left( \dprod\limits_{t=-\infty }^{-1}\left(
\Gamma _{-}\left( 1\right) q^{L_{0}}\right) \right) |0>,  \label{psi-}
\end{equation}%
together with%
\begin{equation}
<0|\Psi _{+}\left( 1\right) =<0|\left( \dprod\limits_{t=0}^{\infty }\left(
q^{L_{0}}\Gamma _{+}\left( 1\right) \right) \right) .  \label{psc}
\end{equation}%
Notice in passing that in eqs(\ref{psi-}-\ref{psc}), the products $%
\dprod\limits_{t=-\infty }^{-1}\left( ...\right) $\ and $\dprod%
\limits_{t=0}^{\infty }\left( ...\right) $\ are taken over diagonal slices
of the 3d partitions. These products are typical ones in the transfer matrix
method where a 3d partition is thought of as a bound state from the slice at 
$t=-\infty $\ (in-state) to the slice at $t=+\infty $\ (out-state). The
action by the operator $\Gamma _{-}(1)$\ allows to generate all possible 2d-
partitions on a given diagonal slice as shown on eq(\ref{222}). The relation
(\ref{sa}) permits to move from a slice to an other by creating all possible
partitions interlacing with the partitions in the previous slice. \newline
Therefore, using eq(\ref{sa}) and $q^{L_{0}}|0>=|0>$, the states (\ref{psi-}-%
\ref{psc}) can be rewritten as 
\begin{equation}
\Psi _{-}\left( 1\right) |0>=\left( \dprod\limits_{k=0}^{\infty }\Gamma
_{-}\left( q^{k}\right) \right) |0>,
\end{equation}%
and similar relation for $<0|\Psi _{+}\left( 1\right) $. We deduce from this
relation the two following:\newline
(\textbf{i}) 3d partitions can be realized in terms of an infinite 2d ones.%
\newline
(\textbf{ii}) the generating functional of 3d partitions are captured by the
local vertex operators\footnote{%
It should be noted that a 3d partition is a collection of Young diagrams.
However, an arbitrary collection of Young diagrams do not correspond to a 3d
partition.}%
\begin{equation}
\Psi _{-}\left( 1\right) =\lim_{s\rightarrow \infty }\left(
\dprod\limits_{k=0}^{s}\Gamma _{-}\left( q^{k}\right) \right) q^{sL_{0}}%
\text{ ,}  \label{psi1}
\end{equation}%
and its dual $\Psi _{+}\left( 1\right) $. These $\Psi _{\pm }$ operators
will be denoted later as 
\begin{equation}
\Psi _{\pm }\left( z\right) =\Gamma _{\pm }^{\left( 2\right) }\left(
z\right) ,
\end{equation}%
but to keep the notations simpler, we will momentary use $\Psi _{\pm }\left(
z\right) $ and come later to the $\Gamma _{\pm }^{\left( 2\right) }$ when we
consider \emph{p- dimensional} generalization.

\subsection{More on 3d generating function}

\qquad The partition function $Z_{3d}$ generating 3d- generalized Young
diagrams is given, in the transfer matrix language, as follows 
\begin{equation}
Z_{3d}=\left\langle 0|\left( \dprod\limits_{t=0}^{\infty }q^{L_{0}}\Gamma
_{+}(1)\right) q^{L_{0}}\left( \dprod\limits_{t=-\infty }^{-1}\Gamma
_{-}(1)q^{L_{0}}\right) |0\right\rangle .  \label{z3}
\end{equation}%
Splitting $q^{L_{0}}$ as $q^{\frac{L_{0}}{2}}q^{\frac{L_{0}}{2}}$ and
commuting each of the operators $q^{\frac{L_{0}}{2}}$ to the left and the
other to the right by using eq(\ref{96}), we get%
\begin{equation}
Z_{3d}=\left\langle 0|\text{ }\dprod\limits_{t=0}^{\infty }\Gamma _{+}\left(
q^{-t-\frac{1}{2}}\right) \dprod\limits_{l=1}^{\infty }\Gamma _{-}\left(
q^{l-\frac{1}{2}}\right) \text{ }|0\right\rangle .
\end{equation}%
Then commuting the $\Gamma _{-}$'s to the left of $\Gamma _{+}$, we obtain%
\begin{equation}
Z_{3d}=\left( \dprod\limits_{l=0}^{\infty }\left[ \dprod\limits_{j=1}^{%
\infty }\left( \frac{1}{\left( 1-q^{j+l}\right) }\right) \right] \right) 
\text{ },  \label{ck}
\end{equation}%
By setting $j+l=k$, we can bring this relation to%
\begin{equation}
Z_{3d}=\left( \dprod\limits_{k=1}^{\infty }\left[ \dprod\limits_{j=1}^{k}%
\left( \frac{1}{\left( 1-q^{k}\right) }\right) \right] \right) \text{ ,}
\end{equation}%
and then to 
\begin{equation}
Z_{3d}=\dprod\limits_{k=1}^{\infty }\left( \frac{1}{\left( 1-q^{k}\right)
^{k}}\right) \text{ },
\end{equation}%
which is precisely the usual form of the 3d- MacMahon function. Before
proceding ahead notice the four following: \newline
\textbf{(1) }$Z_{3d}$\textbf{\ as a \emph{free} CFT}$_{2}$\textbf{\ with
central charge }$\mathbf{c\rightarrow }\infty $ \newline
The expression (\ref{ck}) of $Z_{3d}$ is very suggestive. Setting%
\begin{equation}
Z_{2d}^{\left( l\right) }=\dprod\limits_{j=1}^{\infty }\left( \frac{1}{%
\left( 1-Q_{l}q^{j}\right) }\right) \text{ },\qquad Q_{l}=q^{l}\text{ },
\end{equation}%
which, roughly, describes a partition function $Z_{2d}$, we could then
rewrite eq(\ref{ck}) like%
\begin{equation}
Z_{3d}\text{ }\sim \text{ }\dprod\limits_{l=0}^{\infty }Z_{2d}^{\left(
l\right) }\text{ }.
\end{equation}%
Seen that each $Z_{2d}^{\left( l\right) }$ is associated with a $c=1$ \emph{%
free} CFT$_{2}$ representation, it follows that $Z_{3d}$ could be
interpreted as the partition function of a \emph{free} CFT$_{2}$
representation with $c\rightarrow \infty $. In section 6, we develop an
alternative interpretation of $Z_{3d}$ using correlation of $c=1$ \emph{q-
deformed} vertex operators.\newline
(\textbf{2}) \textbf{Vertex operators} $\Psi _{\pm }\left( z\right) $: \emph{%
Level 2}. \newline
Using eqs(\ref{ga}-\ref{psi-}), it is not difficult to check that $\Psi
_{\pm }\left( z\right) $ is also a local vertex operator whose explicit
expression in terms of the $J_{\pm n}$\ modes, reads as,%
\begin{eqnarray}
\Psi _{-}\left( z\right) &=&\exp \left( i\sum_{n>0}\frac{1}{n}\left( \frac{%
z^{n}}{1-q^{n}}\right) J_{-n}\right) \text{ },  \notag \\
\Psi _{+}\left( z\right) &=&\exp \left( -i\sum_{n>0}\frac{1}{n}\left( \frac{%
z^{-n}}{1-q^{n}}\right) J_{n}\right) \text{ }.
\end{eqnarray}%
The explicit derivation of these relations is given in appendix A, eq(\ref%
{b2}). Notice that:\newline
\textbf{(i)} $\Gamma _{\pm }\left( z\right) $ and $\Psi _{\pm }\left(
z\right) $ are related by the mapping%
\begin{equation}
z^{\pm n}\qquad \rightarrow \qquad \frac{z^{\pm n}}{1-q^{n}}\text{ },
\end{equation}%
Since for $q\rightarrow 0$, $\Gamma _{\pm }\left( z\right) $ and $\Psi _{\pm
}\left( z\right) $ coincide, it follows that the operators $\Psi _{\pm
}\left( z\right) $ can be interpreted as a q- deformation $\Gamma _{\pm
}\left( z\right) $.\newline
\textbf{(ii)} $\Gamma _{\pm }\left( z\right) $ and $\Psi _{\pm }\left(
z\right) $ share most of the basic quantum properties since both of them
involve the same Kac-Moody mode operators $J_{\pm n}$, \newline
\textbf{(3) Translations}\newline
The operator $q^{L_{0}}$ acts also as a translation operator on $\Psi _{\pm
}\left( z\right) $ in the same manner like for $\Gamma _{\pm }\left(
z\right) $.%
\begin{equation}
q^{L_{0}}\Psi _{\pm }\left( z\right) q^{-L_{0}}=\Psi _{\pm }\left( qz\right) 
\text{ },  \label{tr3}
\end{equation}%
This property allows us to define 4d- generalization from 3d one in quite
similar manner as we have done in going from 2d to 3d. We will come back to
this feature later.\newline
\textbf{(4) }$Z_{3d}$\textbf{\ as a "2- point correlation" function}.\newline
Using $\Psi _{\pm }\left( 1\right) $ vertex operators, the partition
function $Z_{3d}$ can be put in the simplest form%
\begin{equation}
Z_{3d}=\left\langle 0|\text{ }\Psi _{+}\left( 1\right) q^{L_{0}}\Psi
_{-}\left( 1\right) |0\right\rangle \text{ }.
\end{equation}%
By help of the identity $q^{L_{0}}\Psi _{-}\left( 1\right) q^{-L_{0}}=\Psi
_{-}\left( q\right) $ eq(\ref{tr3}), we also have%
\begin{equation}
Z_{3d}=\left\langle 0|\Psi _{+}\left( 1\right) \Psi _{-}\left( q\right)
|0\right\rangle \text{ ,}  \label{3d}
\end{equation}%
where $Z_{3d}$ appears as just the 2- point correlation function of the 
\emph{level 2} vertex operators $\Psi _{+}\left( 1\right) $ and $\Psi
_{-}\left( q\right) $. It happens that eq(\ref{3d}) is not the unique way to
define Z$_{3d}$. Let us comment briefly aspects of this issue; general
results will be given in sections 5 and 6.\newline
\textbf{(i)} Eq(\ref{3d}) can be also expressed as follows%
\begin{equation}
Z_{3d}=\left\langle 0|\Gamma _{+}\left( 1\right) q^{L_{0}}\left(
\dprod\limits_{t=-\infty }^{-1}\Psi _{-}\left( 1\right) q^{L_{0}}\right)
|0\right\rangle .
\end{equation}%
This expression will allow us to get the definition of higher dimensional
generalizations of MacMahon function; see eq(\ref{z3p}). This relation can
be put in the simple form%
\begin{equation}
Z_{3d}=\left\langle 0|\Gamma _{+}\left( 1\right) q^{L_{0}}\Omega _{-}\left(
1\right) |0\right\rangle \text{ ,}
\end{equation}%
or equivalently%
\begin{equation}
Z_{3d}=\left\langle 0|\Gamma _{+}\left( 1\right) \Omega _{-}\left( q\right)
|0\right\rangle \text{ },  \label{om-}
\end{equation}%
where we have set%
\begin{equation}
\Omega _{-}\left( 1\right) =\lim_{s\rightarrow \infty }\left(
\dprod\limits_{t=0}^{s}\Psi _{-}\left( q^{k}\right) \right) q^{sL_{0}}\text{
,}  \label{omega}
\end{equation}%
This local vertex operator should be thought of as the \emph{level 3 }of the
hierarchy we have refereed to earlier; i.e.%
\begin{equation}
\Omega _{\pm }\left( 1\right) =\Gamma _{\pm }^{\left( 3\right) }\text{ ,}
\end{equation}%
(\textbf{ii}) Along with the two representations (\ref{3d}) and (\ref{om-})
the partition function $Z_{3d}$ can be expressed as well like 
\begin{equation}
Z_{3d}=\left\langle 0|\Omega _{+}\left( \frac{1}{q}\right) \Gamma _{-}\left(
1\right) |0\right\rangle \text{ ,}
\end{equation}%
where we have used the correlation of $\Omega _{+}\left( x\right) $ and $%
\Gamma _{-}\left( y\right) $ rather than $\Gamma _{+}\left( x\right) $ and $%
\Omega _{-}\left( y\right) $. \newline
(\textbf{iii}) The diversity in expressing $Z_{3d}$ as a 2- point
correlation function, let us suspect that $Z_{3d}$ could be expressed as a
more basic objects. In exploring this idea, we have found that the adequate
interpretation of $Z_{3d}\left( q\right) $ is as a special 4- point
correlation function%
\begin{equation}
Z_{3d}\left( q\right) =\mathcal{G}_{4}\left( x_{0}\text{,}%
x_{1},x_{2},x_{3}\right) \text{ },
\end{equation}%
of vertex operators $\mathcal{O}_{j}\left( x_{j}\right) $ involving
different $\Gamma _{\pm }^{\left( p\right) }$ levels,%
\begin{equation}
\mathcal{G}_{4}=\left\langle 0|\mathcal{O}_{0}\left( x_{0}\right) \mathcal{O}%
_{1}\left( x_{1}\right) \mathcal{O}_{2}\left( x_{2}\right) \mathcal{O}%
_{3}\left( x_{3}\right) |0\right\rangle \text{ }.  \label{g4}
\end{equation}%
To fix the ideas keep in mind the two following: \newline
($\mathbf{\alpha }$) the vertex operator $\mathcal{O}_{0}\left( x_{0}\right) 
$ stands for $\Gamma _{+}\left( 1\right) $ and the other operators will be
explicitly given in section 6; see eqs (\ref{pg}). \newline
($\mathbf{\beta }$) the observed diversity in defining $Z_{3d}\left(
q\right) $ corresponds just to decomposing (\ref{g4}) by using Wick theorem
combined with eqs(\ref{1}-\ref{2}).

\section{Extension to 4d and 5d}

\qquad We first show that the leading terms of the generalized MacMahon
function can be realized as 2- point functions of some vertex operators of $%
c=1$ 2d conformal field theory. \newline
Then, we use this feature to derive the general formula for $G_{d}\left(
q\right) $. In section 5, we consider the interpretation of $G_{d}\left(
q\right) $ as $\left( d+1\right) $- points correlation function $\mathcal{G}%
_{4}\left( x_{0}\text{,}x_{1},\ldots ,x_{d}\right) $ involving vertex
operators $\mathcal{O}_{j}\left( x_{j}\right) $ as in eq(\ref{re}).

\subsection{ $Z_{1d}$ and $Z_{2d}$\ as 2-point functions}

\qquad Before studying 4d and 5d generalizations, it is interesting to start
by revisiting the 1d and 2d cases. This is an important thing for getting
the full picture on the conjectured MacMahon function $G_{d}$.\newline
We start by noting the two following: \newline
\textbf{(1)} Recall that the 1d- MacMahon function corresponds to,%
\begin{equation}
Z_{1d}=\frac{1}{1-q}.
\end{equation}%
This function can be \emph{exactly} interpreted as the two- point correlation%
\footnote{$Z_{pd}$ is the MacMahon function $G_{pd}$; it should'nt be
confused with its interpretation as $\left( p+1\right) $- points correlation
function $\mathcal{G}_{p+1}=\mathcal{G}\left( x_{0},x_{1},...,x_{p}\right) $
to be studied in section 6; see also eqs(\ref{10}).} 
\begin{equation}
Z_{1d}=\mathcal{G}_{2}=\mathcal{G}_{2}\left( z_{0},z_{1}\right) ,
\end{equation}%
of the vertex operators $\Gamma _{+}\left( 1\right) $ and $\Gamma _{-}\left(
q\right) $ as shown below,%
\begin{equation}
Z_{1d}=\left\langle 0|\Gamma _{+}\left( 1\right) \Gamma _{-}\left( q\right)
|0\right\rangle .  \label{1d}
\end{equation}%
This relation describes just the bosonization of eq(\ref{1z}) and can be
rewritten, by using the hamiltonian $L_{0}$, as follows,%
\begin{equation}
Z_{1d}=\left\langle 0|\text{ }\Gamma _{+}\left( 1\right) q^{L_{0}}\Gamma
_{-}\left( 1\right) |0\right\rangle  \label{z1}
\end{equation}%
\textbf{(2)} A quite similar interpretation can be also given for 2d-
MacMahon function,%
\begin{equation}
Z_{2d}=\dprod\limits_{k\geq 1}\left( 1-q^{k}\right) ^{-1}
\end{equation}%
This relation can be expressed in terms of operators vertex as follows,%
\begin{equation}
Z_{2d}=\left\langle 0|\Gamma _{+}\left( 1\right) q^{L_{0}}\left(
\dprod\limits_{k\geq 1}\Gamma _{-}\left( 1\right) q^{L_{0}}\right)
|0\right\rangle ,  \label{z2}
\end{equation}%
Indeed using (\ref{sa}), we can bring it to%
\begin{equation}
Z_{2d}=<0|\Gamma _{+}(1)\left( \dprod\limits_{k\geq 1}\Gamma
_{-}(q^{k})\right) \left\vert 0\right\rangle
\end{equation}%
Then moving the\ operators $\Gamma _{-}\left( q^{k}\right) $\ to the left
and $\Gamma _{+}\left( 1\right) $\ to the right by using eqs(\ref{96}), we
get the desired result.\newline
Notice that using eq(\ref{psi-}), we learn that $Z_{2d}$ can be also defined
as two- point correlation function as follows%
\begin{equation}
Z_{2d}=<0|\Gamma _{+}(1)\Psi _{-}(q)\left\vert 0\right\rangle .  \label{z21}
\end{equation}%
This relation involves the correlation of two vertex operators$\ $of
different levels namely\textrm{\ }$\Gamma _{+}$ (\emph{level 1}) and $\Psi
_{-}(q)$ (\emph{level 2}). Remark also that though $\Gamma _{+}$ and $\Gamma
_{-}$ do not appear on equal footing in eq(\ref{z2}-\ref{z21}), positivity
of $Z_{2d}$ is ensured because $\Gamma _{+}$ and $\Gamma _{-}$ are positive
defined operators. Notice moreover that we also have%
\begin{equation}
Z_{2d}=\mathcal{G}_{3}=\mathcal{G}_{3}\left( z_{0},z_{1},z_{3}\right) ,
\end{equation}%
but this feature will be discussed later on once we give the derivation
proof of the conjectured MacMahon function G$_{d}$.

\subsection{$Z_{pd}$ derivation for $p=4,5$}

\qquad The property that $Z_{1d}$ (\ref{1d}), $Z_{2d}$ (\ref{z21}) and $%
Z_{3d}$ (\ref{3d}) can be all of them interpreted as 2- point correlation
functions of some given vertex operators is very remarkable. It happens in
fact that this feature is a more general property valid also for higher
dimensional generalizations. Let us describe this feature here for the 4d
and 5d cases.

Motivated by the above analysis, 4d- and 5d- generalizations of the MacMahon
function can be then defined as well as 2- point correlation functions of
some local operators as follows%
\begin{eqnarray}
Z_{4d} &=&\left\langle 0|\Psi _{+}\left( 1\right) \Omega _{-}\left( q\right)
|0\right\rangle ,  \label{77} \\
Z_{5d} &=&\left\langle 0|\Omega _{+}\left( 1\right) \Omega _{-}\left(
q\right) |0\right\rangle ,  \notag
\end{eqnarray}%
where $\Omega _{\pm }\left( q\right) $ are vertex operators of some
hierarchy level (\emph{level 3}) which remain to be specified. These
relations have been motivated by the following,%
\begin{eqnarray}
Z_{2d} &=&\left\langle 0|\Gamma _{+}\left( 1\right) \Psi _{-}\left( q\right)
|0\right\rangle ,  \notag \\
Z_{3d} &=&\left\langle 0|\Psi _{+}\left( 1\right) \Psi _{-}\left( q\right)
|0\right\rangle ,
\end{eqnarray}%
and also%
\begin{eqnarray}
Z_{0d} &=&\left\langle 0|I_{id}\left( 1\right) \Gamma _{-}\left( q\right)
|0\right\rangle ,  \notag \\
Z_{1d} &=&\left\langle 0|\Gamma _{+}\left( 1\right) \Gamma _{-}\left(
q\right) |0\right\rangle ,
\end{eqnarray}%
where $I_{id}$ stands for the identity operator (of level zero). To get the $%
\Omega _{\pm }\left( z\right) $ operators, we require:\newline
\textbf{(i)} The $\Omega _{+}\left( z\right) $ and $\Omega _{-}\left(
z\right) $ are local CFT$_{2}$ vertex operators that should obey%
\begin{eqnarray}
\Omega _{-}\left( x\right) \Omega _{-}\left( y\right) &=&\Omega _{-}\left(
y\right) \Omega _{-}\left( y\right) ,  \notag \\
\Omega _{-}\left( x\right) \Psi _{-}\left( y\right) &=&\Psi _{-}\left(
y\right) \Omega _{-}\left( y\right) , \\
\Omega _{-}\left( x\right) \Gamma _{-}\left( y\right) &=&\Gamma _{-}\left(
y\right) \Omega _{-}\left( y\right) ,  \notag \\
\Omega _{-}\left( 0\right) &=&1,  \notag
\end{eqnarray}%
and similar relations for $\Omega _{+}\left( x\right) $.\newline
\textbf{(ii)} We should also have 
\begin{equation}
\Omega _{-}\left( q\right) =q^{L_{0}}\Omega _{-}\left( 1\right) q^{-L_{0}}
\label{tps}
\end{equation}%
so that 
\begin{eqnarray}
Z_{4d} &=&\left\langle 0|\Psi _{+}\left( 1\right) q^{L_{0}}\Omega _{-}\left(
1\right) |0\right\rangle ,  \notag \\
Z_{5d} &=&\left\langle 0|\Omega _{+}\left( 1\right) q^{L_{0}}\Omega
_{-}\left( 1\right) |0\right\rangle ,  \label{z5d}
\end{eqnarray}%
in analogy with the transfer matrix method used previously.\newline
\textbf{(iii)} We impose the commutation relations 
\begin{eqnarray}
\Psi _{+}\left( 1\right) \Omega _{-}\left( q\right) &=&G_{4}\left( q\right)
\Omega _{-}\left( q\right) \Psi _{+}\left( 1\right) ,  \notag \\
\Omega _{+}\left( 1\right) \Omega _{-}\left( q\right) &=&G_{5}\left(
q\right) \Omega _{-}\left( q\right) \Omega _{+}\left( 1\right) ,
\end{eqnarray}%
where $G_{4}\left( q\right) $ and $G_{5}\left( q\right) $ stand for the 4d-
and 5d- generalized MacMahon functions given by, 
\begin{eqnarray}
G_{4}\left( q\right) &=&\dprod\limits_{k=1}^{\infty }\left[ \left( \frac{1}{%
1-q^{k}}\right) ^{\frac{\left( k+1\right) !}{\left( k-1\right) !2!}}\right] ,
\notag \\
G_{5}\left( q\right) &=&\dprod\limits_{k=1}^{\infty }\left[ \left( \frac{1}{%
1-q^{k}}\right) ^{\frac{\left( k+2\right) !}{\left( k-1\right) !3!}}\right] .
\label{4g}
\end{eqnarray}%
A solution of these constraint relations is given by%
\begin{eqnarray}
\Omega _{-}\left( 1\right) &=&\left( \dprod\limits_{t_{2}=-\infty
}^{-1}\left( \dprod\limits_{t_{1}=-\infty }^{-1}\left( \Gamma
_{-}(1)q^{L_{0}}\right) \right) q^{L_{0}}\right) ,  \notag \\
\Omega _{+}\left( 1\right) &=&\left( \dprod\limits_{t_{2}=0}^{\infty
}q^{L_{0}}\left( \dprod\limits_{t_{1}=0}^{\infty }\left( q^{L_{0}}\Gamma
_{+}(1)\right) \right) \right) ,
\end{eqnarray}%
or equivalently like%
\begin{eqnarray}
\Omega _{-}\left( 1\right) &=&\left( \dprod\limits_{t=-\infty }^{-1}\left(
\Psi _{-}\left( 1\right) \right) q^{L_{0}}\right) ,  \notag \\
\Omega _{+}\left( 1\right) &=&\left( \dprod\limits_{t=0}^{\infty
}q^{L_{0}}\Psi _{+}(1)\right) .  \label{wm}
\end{eqnarray}%
To check that these relations solve indeed the above constraint eqs, let us
give some explicit details.

\textbf{4d case}:\newline
First consider the 4d- partition function $Z_{4d}$ expressed in (\ref{z5d})
which we rewrite by substituting (\ref{wm}) as follows,%
\begin{equation}
Z_{4d}=\left\langle 0|\Psi _{+}\left( 1\right) q^{L_{0}}\left(
\dprod\limits_{t=-\infty }^{-1}\Psi _{-}\left( 1\right) q^{L_{0}}\right)
|0\right\rangle .
\end{equation}%
By help of eq(\ref{tps}), it reads also like%
\begin{equation}
Z_{4d}=\left\langle 0|\Psi _{+}\left( 1\right) \left(
\dprod\limits_{l=1}^{\infty }\Psi _{-}\left( q^{l}\right) \right)
|0\right\rangle .
\end{equation}%
Then commuting $\Psi _{-}\left( q^{l}\right) $ to the left by using the
identity%
\begin{equation}
\Psi _{+}\left( 1\right) \Psi _{-}\left( x\right) =\left[ \dprod%
\limits_{k=1}^{\infty }\left( \frac{1}{\left( 1-xq^{k-1}\right) ^{k}}\right) %
\right] \Psi _{-}\left( x\right) \Psi _{+}\left( 1\right) ,\qquad x<1,
\end{equation}%
see also appendix A eqs(\ref{xy}) for general case, we get%
\begin{eqnarray}
Z_{4d} &=&\dprod\limits_{k=1}^{\infty }\dprod\limits_{l=1}^{\infty }\left( 
\frac{1}{\left( 1-q^{l+k-1}\right) ^{k}}\right)  \notag \\
&=&\dprod\limits_{s=1}^{\infty }\dprod\limits_{k=1}^{s}\left( \frac{1}{%
\left( 1-q^{s}\right) ^{k}}\right)
\end{eqnarray}%
which, up on using $\sum\limits_{k=1}^{s}k=\frac{s\left( s+1\right) }{2}$,
can be also put in the form%
\begin{equation}
Z_{4d}=\dprod\limits_{s=1}^{\infty }\left( \frac{1}{\left( 1-q^{s}\right) ^{%
\frac{s\left( s+1\right) }{2}}}\right) ,
\end{equation}%
that should be compared with eq(\ref{4g}). Notice that like for $Z_{3d}$,
the 4d partition function can be expressed in different, but equivalent,
ways: We have the results%
\begin{equation}
Z_{4d}=\left\{ 
\begin{array}{c}
\left\langle 0|\Psi _{+}\left( 1\right) \Omega _{-}\left( q\right)
|0\right\rangle , \\ 
\left\langle 0|\Omega _{+}\left( \frac{1}{q}\right) \Psi _{-}\left( 1\right)
|0\right\rangle , \\ 
\left\langle 0|\Gamma _{+}\left( 1\right) \Upsilon _{-}\left( q\right)
|0\right\rangle , \\ 
\left\langle 0|\Upsilon _{+}\left( \frac{1}{q}\right) \Gamma _{-}\left(
1\right) |0\right\rangle ,%
\end{array}%
\right.
\end{equation}%
where we have set%
\begin{equation}
\Upsilon _{-}\left( q\right) =\left( \dprod\limits_{k=1}^{\infty }\Omega
_{-}\left( q^{k}\right) \right)
\end{equation}%
which should be thought of as $\Upsilon _{-}=\Gamma _{-}^{\left( 4\right) }$.

\textbf{5d case}\newline
Similarly, we have for the 5d- generalization (\ref{z5d}),%
\begin{equation}
Z_{5d}=\left\langle 0|\left( \dprod\limits_{t=0}^{\infty }q^{L_{0}}\Psi
_{+}(1)\right) q^{L_{0}}\Omega _{-}\left( 1\right) |0\right\rangle ,
\end{equation}%
where we have substituted $\Omega _{+}$ in terms of product of $\Psi _{+}$ (%
\ref{wm}). Next using the fact that $q^{L_{0}}$ acts as a translation
operator, we can put $Z_{5d}$ as follows%
\begin{equation}
Z_{5d}=\left\langle 0|\left( \dprod\limits_{l=1}^{\infty }\Psi
_{+}(q^{-l})\right) \Omega _{-}\left( 1\right) |0\right\rangle .
\end{equation}%
Then using the identity 
\begin{equation}
\Psi _{+}(\frac{1}{x})\Omega _{-}\left( 1\right) =\left[ \dprod%
\limits_{s=1}^{\infty }\left( \frac{1}{1-xq^{s}}\right) ^{\frac{s\left(
s+1\right) }{2}}\right] \Omega _{-}\left( 1\right) \Psi _{+}(\frac{1}{x}),
\end{equation}%
we obtain%
\begin{equation}
Z_{5d}=\dprod\limits_{s=1}^{\infty }\dprod\limits_{l=1}^{\infty }\left[
\left( \frac{1}{1-q^{l+s}}\right) ^{\frac{s\left( s+1\right) }{2}}\right] .
\end{equation}%
The next step is to put it in the form%
\begin{equation}
Z_{5d}=\dprod\limits_{k=1}^{\infty }\dprod\limits_{s=1}^{k}\left[ \left( 
\frac{1}{1-q^{k}}\right) ^{\frac{s\left( s+1\right) }{2}}\right] ,
\end{equation}%
which gives%
\begin{equation}
Z_{5d}=\dprod\limits_{k=1}^{\infty }\left[ \left( \frac{1}{1-q^{k}}\right) ^{%
\frac{k\left( k+1\right) \left( k+2\right) }{6}}\right] .
\end{equation}%
In getting this relation, we have used the identity%
\begin{equation}
\sum_{s=1}^{k}\frac{s\left( s+1\right) }{2}=\frac{k\left( k+1\right) \left(
k+2\right) }{6},
\end{equation}%
proved in appendix B. Here also we have different, but equivalent, ways to
define $Z_{5d}$. Later on, we will give the exact numbers of ways for
generic $Z_{pd}$.

\section{Result by induction}

\qquad First notice that the expression (\ref{3d}) of the partition function
Z$_{3d}$ can be also put in the form%
\begin{equation}
Z_{3d}=\left\langle 0|\Gamma _{+}\left( 1\right) q^{L_{0}}\left(
\dprod\limits_{t_{2}=-\infty }^{-1}\left( \Psi _{-}\left( 1\right)
q^{L_{0}}\right) \right) |0\right\rangle
\end{equation}%
or equivalently by using eq(\ref{psi-}), like%
\begin{equation}
Z_{3d}=\left\langle 0|\Gamma _{+}\left( 1\right) q^{L_{0}}\left(
\dprod\limits_{t_{2}=-\infty }^{-1}\left[ \left(
\dprod\limits_{t_{1}=-\infty }^{-1}\left( \Gamma _{-}\left( q\right)
q^{L_{0}}\right) \right) q^{L_{0}}\right] \right) |0\right\rangle .
\label{z3p}
\end{equation}%
This relation as well as eqs(\ref{z1}-\ref{z2}) suggest us the structure of
the p-dimensional partition function $Z_{pd}$ in terms of CFT$_{\mathrm{2}}$%
's vertex operators $\Gamma _{\pm }$. For doing so, we need to introduce the
following hierarchy of local vertex operators%
\begin{equation}
\Gamma _{-}^{\left( n+1\right) }\left( 1\right) =\left(
\dprod\limits_{t_{n}=-\infty }^{-1}\cdots \dprod\limits_{t_{2}=-\infty
}^{-1} \left[ \left( \dprod\limits_{t_{1}=-\infty }^{-1}\left( \Gamma
_{-}\left( 1\right) q^{L_{0}}\right) \right) q^{L_{0}}\right] \cdots
q^{L_{0}}\right)  \label{gan}
\end{equation}%
for $n\geq 1$, together with%
\begin{equation}
\Gamma _{-}^{\left( 0\right) }=I_{id},\qquad \Gamma _{-}^{\left( 1\right)
}\left( z\right) =\Gamma _{-}\left( z\right) .
\end{equation}%
Eq(\ref{gan}) can be also defined as follows,%
\begin{equation}
\Gamma _{-}^{\left( n+1\right) }\left( 1\right) =\dprod\limits_{t=-\infty
}^{-1}\left( \Gamma _{-}^{\left( n\right) }\left( 1\right) q^{L_{0}}\right)
,\qquad n\geq 1.  \label{gap}
\end{equation}%
A similar relation can be written down for $\Gamma _{+}^{\left( n+1\right)
}\left( 1\right) $. The $\Gamma _{-}^{\left( p\right) }$, referred to as the 
\emph{level p} vertex operator, obey quite similar relations that the ones
associated to $\Gamma _{-}\left( 1\right) $, in particular 
\begin{equation}
\Gamma _{-}^{\left( p\right) }\left( q\right) =q^{L_{0}}\Gamma _{-}^{\left(
p\right) }\left( 1\right) q^{-L_{0}},\qquad p\geq 0.
\end{equation}%
More details, concerning these high level operators, are presented in
Appendix A.\newline
Based on the preceding results realized for lower dimensions, it follows
that the $p$- dimensional partition functions Z$_{pd}$ can be defined as, 
\begin{equation}
Z_{pd}=\left\langle 0|\Gamma _{+}\left( 1\right) \Gamma _{-}^{\left(
p\right) }\left( q\right) |0\right\rangle ,\qquad p\geq 0.  \label{pd}
\end{equation}%
This relation, which has been explicitly checked for $p=0,1,2,3,4$ and $5$,
reads also as%
\begin{equation}
Z_{pd}=\left\langle 0|\Gamma _{+}\left( 1\right) q^{L_{0}}\Gamma
_{-}^{\left( p\right) }\left( 1\right) |0\right\rangle .  \label{gn}
\end{equation}%
Commuting $\Gamma _{-}^{\left( p\right) }\left( q\right) $ to the left of $%
\Gamma _{+}\left( 1\right) $, we can show by induction that for $p\geq 2$ 
\begin{equation}
Z_{pd}=\dprod\limits_{k=1}^{\infty }\left[ \left( \frac{1}{1-q^{k}}\right) ^{%
\frac{\left( k+p-3\right) !}{\left( k-1\right) !\left( p-2\right) !}}\right]
,  \label{zp}
\end{equation}%
\emph{Proof by induction}\textrm{:} \newline
We suppose that eq(\ref{zp}) holds for \emph{level p}; then prove that it
holds as well for \emph{level }$\left( p+1\right) $; that is, 
\begin{equation}
Z_{\left( p+1\right) d}=\left\langle 0|\Gamma _{+}\left( 1\right) \Gamma
_{-}^{\left( p+1\right) }\left( q\right) |0\right\rangle ,  \label{qz}
\end{equation}%
and find that it is given by%
\begin{equation}
Z_{\left( p+1\right) d}=\dprod\limits_{k=1}^{\infty }\left[ \left( \frac{1}{%
1-q^{k}}\right) ^{\frac{\left( k+p-2\right) !}{\left( k-1\right) !\left(
p-1\right) !}}\right] .
\end{equation}%
Indeed, we start from the definition of $Z_{\left( p+1\right) d}$, 
\begin{equation}
Z_{\left( p+1\right) d}=\left\langle 0|\Gamma _{+}\left( 1\right) \Gamma
_{-}^{\left( p+1\right) }\left( q\right) |0\right\rangle
\end{equation}%
and express it, by using (\ref{gap}), as 
\begin{equation}
Z_{\left( p+1\right) d}=\left\langle 0|\Gamma _{+}\left( 1\right)
q^{L_{0}}\left( \dprod\limits_{t=-\infty }^{-1}\Gamma _{-}^{\left( p\right)
}\left( 1\right) q^{L_{0}}\right) |0\right\rangle
\end{equation}%
or equivalently like%
\begin{equation}
Z_{\left( p+1\right) d}=\left\langle 0|\Gamma _{+}\left( 1\right) \left(
\dprod\limits_{l=1}^{\infty }\Gamma _{-}^{\left( p\right) }\left(
q^{l}\right) \right) |0\right\rangle .
\end{equation}%
Then we commute $\Gamma _{-}^{\left( p\right) }\left( q^{l}\right) $ to the
left of $\Gamma _{+}\left( 1\right) $ in eq(\ref{qz}), we get after some
computations, 
\begin{equation}
Z_{\left( p+1\right) d}=\dprod\limits_{l=1}^{\infty
}\dprod\limits_{k=1}^{\infty }\left[ \left( \frac{1}{1-q^{l+k}}\right) ^{%
\frac{\left( k+p-3\right) !}{\left( k-1\right) !\left( p-2\right) !}}\right]
.
\end{equation}%
Setting $s=\left( l+k\right) $, we can rewrite this relation as follows%
\begin{equation}
Z_{\left( p+1\right) d}=\dprod\limits_{s=1}^{\infty }\dprod\limits_{k=1}^{s} 
\left[ \left( \frac{1}{1-q^{s}}\right) ^{\frac{\left( k+p-3\right) !}{\left(
k-1\right) !\left( p-2\right) !}}\right] .
\end{equation}%
At first sight, this expression seems different from the desired result;
however explicit computation leads exactly to the right result; thanks to
the combinatorial identity, 
\begin{equation}
\sum_{k=1}^{s}\frac{\left( k+p-3\right) !}{\left( k-1\right) !\left(
p-2\right) !}=\frac{\left( s+p-2\right) !}{\left( s-1\right) !\left(
p-1\right) !},\quad p\geq 2,  \label{cr}
\end{equation}%
which is \textrm{showed} in appendix B.\newline
These computations give an explicit proof for the derivation of the
expression of generalized MacMahon function. Thanks to the "transfer matrix
method" and to the hierarchy of \emph{level p} vertex operators $\Gamma
_{\pm }^{\left( p\right) }$ eq(\ref{gap}).

\section{$G_{n}\left( q\right) $ as $\left( n+1\right) $-point correlation
function}

\qquad So far we have seen that \emph{n- dimensional} generalization of
MacMahon function $G_{n}\left( q\right) $ with $n\geq 2$, can be interpreted
as 2- point correlation functions of some composite vertex operators.\textrm{%
\ }We have also seen that there are different, but equivalent ways, to
express $G_{n}\left( q\right) $ as 2- point correlation functions. Using $%
\Gamma _{\pm }^{\left( r\right) }$ and $\Gamma _{\pm }^{\left( s\right) }$
vertex operators, one can check that for any positive definite integers r
and s such that $r+s-1=n$, we have,%
\begin{equation}
G_{n}\left( q\right) =<0|\Gamma _{+}^{\left( n-s+1\right) }\left( 1\right)
\Gamma _{-}^{\left( s\right) }\left( q\right) |0>,\qquad 1\leq s\leq
n.\qquad n\geq 1.  \label{div}
\end{equation}%
The $2\left( \frac{r+s-2}{2}\right) +1$ possibilities are all of them equal
to each other. This diversity in defining $G_{n}\left( q\right) $ suggests
us to look for a more refined definition of it. We have found that the
adequate way to define $G_{n}\left( q\right) $ is like a $\left( n+1\right) $%
- point correlation function as given below, 
\begin{equation}
G_{n}\left( q\right) =\mathcal{G}_{n+1}\left( x_{0},x_{1},x_{2}\cdots
,x_{n}\right)
\end{equation}%
with 
\begin{equation}
\mathcal{G}_{n+1}=\left\langle 0|\mathcal{O}_{0}\left( x_{0}\right) \mathcal{%
O}_{1}\left( x_{1}\right) \mathcal{O}_{2}\left( x_{2}\right) \cdots \mathcal{%
O}_{n}\left( x_{n}\right) |0\right\rangle ,  \label{gdn}
\end{equation}%
where the $x_{j}=x_{j}\left( q\right) $ and $\mathcal{O}_{j}\left(
x_{j}\right) $ are some vertex operators that have to be specified. In this
way, the diversity (\ref{div}) appears just as a manifestation of applying
Wick theorem to (\ref{gdn}) for its decomposition in terms of two- points
correlation functions. To fix the ideas, think about $\mathcal{O}_{0}\left(
x_{0}\right) $ as given by%
\begin{equation}
\mathcal{O}_{0}\left( x_{0}\right) =\Gamma _{+}\left( 1\right)
\end{equation}%
and all remaining others as given by vertex operators involving products of $%
\Gamma _{-}\left( y\right) $ only, that is: 
\begin{equation}
\mathcal{O}_{j}\left( x_{j}\right) \sim \dprod \Gamma _{-}\left( y\right)
,\qquad j=1,...,n.
\end{equation}%
Since $\Gamma _{-}\left( y\right) $ and any product of $\Gamma _{-}\left(
y\right) $ has vacuum expectation values equal to one,%
\begin{equation}
\left\langle 0|\Gamma _{-}\left( y\right) |0\right\rangle =1=\left\langle
0|\left( \dprod \Gamma _{-}\left( y\right) \right) |0\right\rangle ,
\end{equation}%
it follows that%
\begin{equation}
\left\langle 0|\dprod\limits_{l=1}^{n}\mathcal{O}_{l}\left( x_{l}\right)
|0\right\rangle =1.  \label{o-}
\end{equation}%
Then, by using Wick theorem $\mathcal{G}_{n+1}$ reduces to%
\begin{equation}
\mathcal{G}_{n+1}=\dprod\limits_{k}\left\langle 0|\mathcal{O}_{0}\left(
x_{0}\right) \mathcal{O}_{k}\left( x_{k}\right) |0\right\rangle ,\qquad
k=1,...,n.
\end{equation}%
Let us build the Green function $\mathcal{G}_{n+1}$ step by step, starting
from eq(\ref{pd}) and using the results obtained above:

\subsection{Leading terms}

\qquad Below, we give the explicit computation of $\mathcal{G}_{n}$ for $%
n=2, $ $3,$ $4,$ $5$\newline
\textbf{(1)} $G_{1}\left( q\right) $ \textbf{as 2- point propagator}.\newline
Comparing%
\begin{equation}
G_{1}\left( q\right) =\left\langle 0|\Gamma _{+}\left( 1\right) \Gamma
_{-}\left( q\right) |0\right\rangle
\end{equation}%
with%
\begin{equation}
\mathcal{G}_{2}=\left\langle 0|\mathcal{O}_{0}\left( x_{0}\right) \mathcal{O}%
_{1}\left( x_{1}\right) |0\right\rangle
\end{equation}%
we get%
\begin{eqnarray}
\mathcal{O}_{0}\left( x_{0}\right) &=&\Gamma _{+}\left( 1\right) ,  \notag \\
\mathcal{O}_{1}\left( x_{1}\right) &=&\Gamma _{-}\left( q\right)  \label{w2}
\end{eqnarray}%
\textbf{(2) }$G_{2}\left( q\right) $ \textbf{as a }3-\textbf{point function}%
\newline
Starting from the expression eq(\ref{pd}) for G$_{2},$%
\begin{equation}
G_{2}\left( q\right) =\left\langle 0|\Gamma _{+}\left( 1\right) \Gamma
_{-}^{\left( 2\right) }\left( q\right) |0\right\rangle ,
\end{equation}%
and using the special property established in appendix A; see eq(\ref{bsy}),%
\begin{equation}
\Gamma _{-}^{\left( 2\right) }\left( q\right) =\Gamma _{-}^{\left( 1\right)
}\left( q\right) \Gamma _{-}^{\left( 2\right) }\left( q^{2}\right)
\label{h2}
\end{equation}%
we can bring $G_{2}\left( q\right) $ into the form%
\begin{equation}
G_{2}\left( q\right) =\left\langle 0|\Gamma _{+}\left( 1\right) \Gamma
_{-}^{\left( 1\right) }\left( q\right) \Gamma _{-}^{\left( 2\right) }\left(
q^{2}\right) |0\right\rangle
\end{equation}%
Then, comparing with%
\begin{equation}
\mathcal{G}_{3}=\left\langle 0|\mathcal{O}_{0}\left( x_{0}\right) \mathcal{O}%
_{1}\left( x_{1}\right) \mathcal{O}_{2}\left( x_{2}\right) |0\right\rangle ,
\end{equation}%
we get, in addition to eqs(\ref{w2}), the following:%
\begin{equation}
\mathcal{O}_{2}\left( x_{2}\right) =\Gamma _{+}^{\left( 2\right) }\left(
q^{2}\right) .  \label{d2}
\end{equation}%
\textbf{(3) }$G_{3}\left( q\right) $ \textbf{as a 4-point function}\newline
We start from the expression of $G_{3}\left( q\right) $,%
\begin{equation}
G_{3}\left( q\right) =\left\langle 0|\Gamma _{+}\left( 1\right) \Gamma
_{-}^{\left( 3\right) }\left( q\right) |0\right\rangle
\end{equation}%
then use the identity,%
\begin{equation}
\Gamma _{-}^{\left( 3\right) }\left( q\right) =\Gamma _{-}^{\left( 2\right)
}\left( q\right) \Gamma _{-}^{\left( 3\right) }\left( q^{2}\right) ,
\label{h3}
\end{equation}%
and substitute $\Gamma _{-}^{\left( 2\right) }\left( q\right) $ by eq(\ref%
{h2}), we get%
\begin{equation}
\Gamma _{-}^{\left( 3\right) }\left( q\right) =\Gamma _{-}^{\left( 1\right)
}\left( q\right) \left( \dprod\limits_{l=1}^{2}\Gamma _{-}^{\left( 2\right)
}\left( q^{2}\right) \right) \Gamma _{-}^{\left( 3\right) }\left(
q^{3}\right) .
\end{equation}%
Comparing with%
\begin{equation}
\mathcal{G}_{3}=\left\langle 0|\mathcal{O}_{0}\left( x_{0}\right) \mathcal{O}%
_{1}\left( x_{1}\right) \mathcal{O}_{2}\left( x_{2}\right) \mathcal{O}%
_{3}\left( x_{3}\right) |0\right\rangle ,
\end{equation}%
we obtain%
\begin{eqnarray}
\mathcal{O}_{0}\left( x_{0}\right) &=&\Gamma _{+}\left( 1\right)  \notag \\
\mathcal{O}_{1}\left( x_{1}\right) &=&\Gamma _{-}^{\left( 1\right) }\left(
q\right)  \notag \\
\mathcal{O}_{2}\left( x_{2}\right) &=&\left( \dprod\limits_{l=1}^{2}\Gamma
_{-}^{\left( 2\right) }\left( q^{2}\right) \right) \\
\mathcal{O}_{3}\left( x_{3}\right) &=&\Gamma _{-}^{\left( 3\right) }\left(
q^{3}\right)  \notag
\end{eqnarray}%
\textbf{(4) }$G_{4}\left( q\right) $ \textbf{as a 5-point function}\newline
Starting from 
\begin{equation}
G_{4}\left( q\right) =\left\langle 0|\Gamma _{+}\left( 1\right) \Gamma
_{-}^{\left( 4\right) }\left( q\right) |0\right\rangle
\end{equation}%
then using the identities,%
\begin{eqnarray}
\Gamma _{-}^{\left( 4\right) }\left( q\right) &=&\Gamma _{-}^{\left(
3\right) }\left( q\right) \Gamma _{-}^{\left( 4\right) }\left( q^{2}\right) ,
\notag \\
\Gamma _{-}^{\left( 4\right) }\left( q^{2}\right) &=&\Gamma _{-}^{\left(
3\right) }\left( q^{2}\right) \Gamma _{-}^{\left( 4\right) }\left(
q^{3}\right) \\
\Gamma _{-}^{\left( 4\right) }\left( q^{3}\right) &=&\Gamma _{-}^{\left(
3\right) }\left( q^{3}\right) \Gamma _{-}^{\left( 4\right) }\left(
q^{4}\right)  \notag
\end{eqnarray}%
we obtain at a first stage%
\begin{equation}
\Gamma _{-}^{\left( 4\right) }\left( q\right) =\Gamma _{-}^{\left( 3\right)
}\left( q\right) \Gamma _{-}^{\left( 3\right) }\left( q^{2}\right) \Gamma
_{-}^{\left( 3\right) }\left( q^{3}\right) \Gamma _{-}^{\left( 4\right)
}\left( q^{4}\right) .  \label{h4}
\end{equation}%
At a second stage, we substitute $\Gamma _{-}^{\left( 3\right) }\left(
q\right) $ and $\Gamma _{-}^{\left( 3\right) }\left( q^{2}\right) $ by eq(%
\ref{h3}), we get%
\begin{eqnarray}
\Gamma _{-}^{\left( 3\right) }\left( q\right) &=&\Gamma _{-}^{\left(
2\right) }\left( q\right) \Gamma _{-}^{\left( 3\right) }\left( q^{2}\right) ,
\notag \\
\Gamma _{-}^{\left( 3\right) }\left( q^{2}\right) &=&\Gamma _{-}^{\left(
2\right) }\left( q^{2}\right) \Gamma _{-}^{\left( 3\right) }\left(
q^{3}\right) .
\end{eqnarray}%
Putting back into eq(\ref{h4}), we obtain%
\begin{equation*}
\Gamma _{-}^{\left( 4\right) }\left( q\right) =\Gamma _{-}^{\left( 2\right)
}\left( q\right) \Gamma _{-}^{\left( 2\right) }\left( q^{2}\right) \Gamma
_{-}^{\left( 2\right) }\left( q^{2}\right) \Gamma _{-}^{\left( 3\right)
}\left( q^{3}\right) \Gamma _{-}^{\left( 3\right) }\left( q^{3}\right)
\Gamma _{-}^{\left( 3\right) }\left( q^{3}\right) \Gamma _{-}^{\left(
4\right) }\left( q^{4}\right) .
\end{equation*}%
Next replacing $\Gamma _{-}^{\left( 2\right) }\left( q\right) $ by eq(\ref%
{h2}), we end with the following, 
\begin{equation}
\Gamma _{-}^{\left( 4\right) }\left( q\right) =\Gamma _{-}^{\left( 1\right)
}\left( q\right) \left( \dprod\limits_{l=1}^{3}\Gamma _{-}^{\left( 2\right)
}\left( q^{2}\right) \right) \left( \dprod\limits_{l=1}^{3}\Gamma
_{-}^{\left( 3\right) }\left( q^{3}\right) \right) \Gamma _{-}^{\left(
4\right) }\left( q^{4}\right) .
\end{equation}%
Comparing with%
\begin{equation}
\mathcal{G}_{5}=\left\langle 0|\mathcal{O}_{0}\left( x_{0}\right) \mathcal{O}%
_{1}\left( x_{1}\right) \mathcal{O}_{2}\left( x_{2}\right) \mathcal{O}%
_{3}\left( x_{3}\right) \mathcal{O}_{4}\left( x_{4}\right) |0\right\rangle ,
\end{equation}%
we obtain%
\begin{eqnarray}
\mathcal{O}_{0}\left( x_{0}\right) &=&\Gamma _{+}\left( 1\right) ,  \notag \\
\mathcal{O}_{1}\left( x_{1}\right) &=&\Gamma _{-}^{\left( 1\right) }\left(
q\right) ,  \notag \\
\mathcal{O}_{2}\left( x_{2}\right) &=&\left( \dprod\limits_{l=1}^{3}\Gamma
_{-}^{\left( 2\right) }\left( q^{2}\right) \right) ,  \notag \\
\mathcal{O}_{3}\left( x_{3}\right) &=&\left( \dprod\limits_{l=1}^{3}\Gamma
_{-}^{\left( 3\right) }\left( q^{3}\right) \right) , \\
\mathcal{O}_{4}\left( x_{4}\right) &=&\Gamma _{-}^{\left( 4\right) }\left(
q^{4}\right) .  \notag
\end{eqnarray}%
With these results on lower values of p, it is straightforward to derive the
generic picture.

\subsection{Generic result}

We start from the expression of 
\begin{equation}
G_{p}\left( q\right) =\left\langle 0|\Gamma _{+}\left( 1\right) \Gamma
_{-}^{\left( p\right) }\left( q\right) |0\right\rangle ,\qquad p\geq 1.
\end{equation}%
Then we use the identity,%
\begin{equation}
\Gamma _{-}^{\left( p\right) }\left( q\right)
=\dprod\limits_{k=0}^{p-1}\left( \dprod\limits_{l_{k}=1}^{p_{k}}\Gamma
_{-}^{\left( k+1\right) }\left( q^{k+1}\right) \right) ,  \label{gm}
\end{equation}%
with%
\begin{equation}
p_{k}=\frac{\left( p-1\right) !}{k!\left( p-k-1\right) !},\qquad 0\leq k\leq
p-1,
\end{equation}%
which is proved in appendix A, eq(\ref{at}), we can bring $G_{p}\left(
q\right) $ into the form,%
\begin{equation}
G_{p}\left( q\right) =\left\langle 0|\Gamma _{+}\left( 1\right)
\dprod\limits_{k=0}^{p-1}\left( \dprod\limits_{l_{k}=1}^{p_{k}}\Gamma
_{-}^{\left( k+1\right) }\left( q^{k+1}\right) \right) |0\right\rangle
,\qquad p\geq 1.  \label{g1}
\end{equation}%
By comparing with, 
\begin{equation}
\mathcal{G}_{p+1}=\left\langle 0|\mathcal{O}_{0}\left( x_{0}\right) \mathcal{%
O}_{1}\left( x_{1}\right) \mathcal{O}_{2}\left( x_{2}\right) \cdots \mathcal{%
O}_{p}\left( x_{p}\right) |0\right\rangle ,  \label{g2}
\end{equation}%
we obtain%
\begin{eqnarray}
\mathcal{O}_{0}\left( x_{0}\right) &=&\Gamma _{+}\left( 1\right) ,\qquad 
\mathcal{O}_{1}\left( x_{1}\right) =\Gamma _{-}^{\left( 1\right) }\left(
q\right) ,  \notag \\
\mathcal{O}_{j}\left( x_{j}\right) &=&\left( \dprod\limits_{l_{j}=1}^{p_{j}}%
\left[ \Gamma _{-}^{\left( j\right) }\left( q^{j}\right) \right] \right)
,\qquad j\geq 2.  \label{pg}
\end{eqnarray}%
For more details on the derivation of this relation, see appendix A: eqs(\ref%
{as})-(\ref{at}). Eqs(\ref{pg}) complete the interpretation of G$_{d}$ as a $%
\left( d+1\right) $- points Green function.

\section{Discussion and Conclusion}

\qquad In this paper, we have given a 2d- conformal field theoretical
derivation of the generalized MacMahon function by using ideas from
"transfer matrix method" and q-deformed QFT$_{2}$. Among our results, we
mention:\newline
\textbf{(1)} The usual vertex operators $\Gamma _{\pm }\left( z\right) $ of
the bosonic $c=1$ conformal field theory appear as the \emph{level one} of
the following hierarchy,%
\begin{equation}
\Gamma _{-}^{\left( p\right) }\left( z\right) \left\vert 0\right\rangle
=\exp \left( \sum_{n=1}^{\infty }\frac{iz^{n}J_{-n}}{n\left( 1-q^{n}\right)
^{p-1}}\right) \left\vert 0\right\rangle ,\qquad p\geq 1
\end{equation}%
where $q=\exp \left( -g_{s}\right) $. These local operators, which coincide
in the limit $q\rightarrow 0$; that is when $g_{s}$ goes to $\infty $, can
be obtained from $\Gamma _{\pm }\left( z\right) $ by making the substitution 
\begin{equation}
z^{n}\qquad \rightarrow \qquad \frac{z^{n}}{\left( 1-q^{n}\right) ^{p-1}}%
,\qquad p\geq 2.
\end{equation}%
The $\Gamma _{-}^{\left( p\right) }$s form then an infinite hierarchy of
q-deformed vertex operators and obey commutation relations quite similar to
those satisfied by the \emph{level one} $\Gamma _{-}^{\left( 1\right)
}\left( z\right) =\Gamma _{-}\left( z\right) $. In particular we have,%
\begin{equation}
\Gamma _{+}^{\left( 1\right) }\left( 1\right) \Gamma _{-}^{\left( p\right)
}\left( q\right) =G_{p}\left( q\right) \Gamma _{-}^{\left( p\right) }\left(
q\right) \Gamma _{+}^{\left( 1\right) }\left( 1\right) ,
\end{equation}%
where $G_{p}\left( q\right) $ is precisely the generalized \emph{p-dimension}
MacMahon function. We also have the following general relation,%
\begin{equation}
\left\langle 0\right\vert \Gamma _{+}^{\left( 1\right) }\left( z_{1}\right)
\Gamma _{-}^{\left( l+1\right) }\left( z_{l}\right) \left\vert
0\right\rangle =\dprod\limits_{k_{l}=0}{\small \cdots }\dprod%
\limits_{k_{1}=0}\left[ \dprod\limits_{k_{0}=0}\left( \frac{1}{\left(
1-q^{k_{0}+k_{1}+\ldots +k_{l}}\frac{z_{l}}{z_{1}}\right) }\right) \right] .
\label{rhs}
\end{equation}%
This relation can be given an interpretation as $l$ copies of $c=\infty $
free CFT$_{2}$ representations. Indeed, setting $\frac{z_{l}}{z_{1}}=q$, $%
q^{k_{0}+k_{1}+\ldots k_{l}}=Q_{\mathbf{k}}q^{k_{0}}$\ with $Q_{\mathbf{k}%
}=q^{k_{1}+\ldots +k_{l}}$ and 
\begin{equation}
Z_{2}\left( Q_{\mathbf{k}},q\right) =\dprod\limits_{k_{0}=0}^{\infty }\left( 
\frac{1}{\left( 1-Q_{\mathbf{k}}q^{k_{0}}\right) }\right) ,\qquad \mathbf{k=}%
\left( k_{1},\ldots ,k_{l}\right) ,
\end{equation}%
we can put the right hand side of (\ref{rhs}) like, 
\begin{equation}
\dprod\limits_{\left( k_{1},\ldots ,k_{l}\right) =\mathbf{0}}^{\mathbf{%
\infty }}Z_{2}\left( Q_{\mathbf{k}},q\right) \text{ }\sim \text{ }\left( %
\left[ Z_{2}\right] ^{\infty }\right) ^{l}.
\end{equation}%
This factorization suggests that, roughly, $G_{l+1}\left( q\right) $ could
be interpreted as given by the product of $l$ copies of infinite products of 
$Z_{2}$. Since from 2d conformal free field theory view, each $Z_{2}^{\infty
}$ copy should be described by a free field CFT$_{2}$ representation with $%
c=\infty $, the $G_{l+1}\left( q\right) $ partition function would then
correspond to a central charge 
\begin{equation}
c=k^{l},
\end{equation}%
with $k\rightarrow \infty $. \newline
\textbf{(2)} Using the above \emph{level p} vertex operators, we have shown
that the \emph{p- dimensional} generalized MacMahon function is given by the
following two- point correlation function%
\begin{equation}
G_{p}\left( q\right) =\left\langle 0\right\vert \Gamma _{+}^{\left( 1\right)
}\left( 1\right) \Gamma _{-}^{\left( p\right) }\left( q\right) \left\vert
0\right\rangle
\end{equation}%
\textbf{(3)} The \emph{level p} vertex operators $\Gamma _{-}^{\left(
p\right) }$ satisfy several remarkable properties, in particular they can be
realized as condensates of vertex operators of \emph{lower levels} as shown
below,%
\begin{equation}
\Gamma _{-}^{\left( p\right) }\left( z\right) =\Gamma _{-}^{\left(
p-1\right) }\left( z\right) \Gamma _{-}^{\left( p\right) }\left( qz\right) ,
\end{equation}%
so that $G_{p}\left( q\right) $ can be defined as a particular $\left(
p+1\right) $-point correlation function as given below,%
\begin{equation}
\mathcal{G}_{p+1}=\left\langle 0|\mathcal{O}_{0}\left( x_{0}\right) \mathcal{%
O}_{1}\left( x_{1}\right) \mathcal{O}_{2}\left( x_{2}\right) \cdots \mathcal{%
O}_{p}\left( x_{p}\right) |0\right\rangle ,
\end{equation}%
where the $\mathcal{O}_{j}\left( x_{j}\right) $ are given by eqs(\ref{g1}-%
\ref{g2}). This correlation function can be expressed in different forms by
using Wick theorem and the property (\ref{o-}).\newline
\textbf{(4)} Based on the field theoretical derivation given in the present
study, we learn that the function $G_{p}\left( q\right) $ with $p\geq 4$
cannot be the generating functional of the \emph{p- dimensional generalized}
Young diagrams. \newline
Recall that for the case $p=3$, solid partitions $\Pi ^{\left( 3\right) }$
extending Young diagrams have generally three boundaries given by 2d
partitions $\lambda ,$ $\mu $ and $\nu $. The typical generating functional
of all possible 3d partitions $\Psi ^{\left( 3\right) }$ with boundaries $%
\partial \left( \Psi ^{\left( 3\right) }\right) =\left( \lambda ,\mu ,\nu
\right) $ is given by the correlation function $C_{\lambda \mu \nu }$ 
\begin{equation}
C_{\lambda \mu \nu }=\left\langle \nu ^{t}|\mathcal{A}_{+}\left( \lambda
\right) \mathcal{A}_{-}\left( \lambda ^{t}\right) |\mu \right\rangle ,
\label{ccc}
\end{equation}%
where $\mathcal{A}_{+}\left( \lambda \right) \mathcal{A}_{-}\left( \lambda
^{t}\right) $ is the transfer matrix operator described previously. For the
simplest case where $\partial \left( \Psi ^{\left( 3\right) }\right) =\left(
\emptyset ,\emptyset ,\emptyset \right) $, the correlation function $%
C_{\emptyset \emptyset \emptyset }$ is precisely the generating functional
of 3d partitions.\newline
For higher values of $p$; say $p=4$, one has 4d generalized Young diagrams $%
\Psi ^{\left( 4\right) }$. This 4d partitions have generally \emph{four} 3-
dimensional\emph{\ boundaries} captured by 3d partitions $\Lambda ^{\left(
3\right) },$ $\Sigma ^{\left( 3\right) }$, $\Upsilon ^{\left( 3\right) }$
and $\Pi ^{\left( 3\right) }$. The typical generating functional of all
possible 4d partitions $\Pi ^{\left( 4\right) }$ with boundaries $\partial
\left( \Psi ^{\left( 4\right) }\right) =\left( \Lambda ^{\left( 3\right)
},\Sigma ^{3},\Upsilon ^{\left( 3\right) },\Pi ^{\left( 3\right) }\right) $
is given by the correlation function $C_{\Lambda \Sigma \Upsilon \Psi }$.
This functional extends (\ref{ccc}) and can be defined as 
\begin{equation}
\left\langle \left\langle \Lambda ^{\left( 3\right) }||\mathcal{A}%
_{-}^{\prime }\left( \Sigma ^{3},\Upsilon ^{\left( 3\right) }\right) 
\mathcal{A}_{+}^{\prime }\left( \Sigma ^{3},\Upsilon ^{\left( 3\right)
}\right) ||\Pi ^{\left( 3\right) }\right\rangle \right\rangle ,
\end{equation}%
where $\mathcal{A}_{-}^{\prime }\left( \Sigma ^{3},\Upsilon ^{\left(
3\right) }\right) \mathcal{A}_{+}^{\prime }\left( \Sigma ^{3},\Upsilon
^{\left( 3\right) }\right) $ is some generalized transfer matrix operator
acting on 3d partition states $||\Pi ^{\left( 3\right) }>>$. It is this
function that would generate the 4d generalized Young diagrams with
boundaries $\Lambda $, $\Sigma $, $\Upsilon $ and $\Pi $. \newline
Moreover, using the fact that 3d partitions $\Pi ^{\left( 3\right) }$ may
themselves be sliced in terms of 2d partitions, one can usually bring the
correlation function $C_{\Lambda \Sigma \Upsilon \Psi }$ to the form, 
\begin{equation}
\left\langle \left\langle \varsigma ,\tau ,\upsilon ||\mathcal{A}%
_{+}^{\prime }\left[ \left( \lambda ,\mu ,\nu \right) ;\left( \zeta ,\eta
,\theta \right) \right] \mathcal{A}_{-}^{\prime }\left[ \left( \lambda
^{t},\mu ^{t},\nu ^{t}\right) ;\left( \zeta ^{t},\eta ^{t},\theta
^{t}\right) \right] ||\alpha ,\beta ,\gamma \right\rangle \right\rangle ,
\end{equation}%
where $\left\vert \left\vert \vartheta ,\sigma ,\varrho \right\rangle
\right\rangle $ is a 3d partition boundary state expressed in terms of 2d
partitions $\left\vert \vartheta \right\rangle \otimes $ $\left\vert \sigma
\right\rangle \otimes $ $\left\vert \varrho \right\rangle $. In the
particular case $\alpha =\beta =\gamma =\emptyset $ and $\varsigma =\tau
=\upsilon =\emptyset $, the correlation function becomes 
\begin{equation}
\left\langle \left\langle \emptyset ,\emptyset ,\emptyset ||\mathcal{A}%
_{+}^{\prime }\left[ \left( \lambda ,\mu ,\nu \right) ;\left( \zeta ,\eta
,\theta \right) \right] \mathcal{A}_{-}^{\prime }\left[ \left( \lambda
^{t},\mu ^{t},\nu ^{t}\right) ;\left( \zeta ^{t},\eta ^{t},\theta
^{t}\right) \right] ||\emptyset ,\emptyset ,\emptyset \right\rangle
\right\rangle .
\end{equation}%
\newline
In the special case $\zeta =\eta =\theta =\lambda =\mu =\nu =\emptyset $,
the above quantity simplifies as%
\begin{equation}
\left\langle \left\langle \emptyset ,\emptyset ,\emptyset ||\mathcal{A}%
_{+}^{\prime }\left[ \emptyset \right] \mathcal{A}_{-}^{\prime }\left[
\emptyset \right] ||\emptyset ,\emptyset ,\emptyset \right\rangle
\right\rangle .
\end{equation}%
\newline
From this general relation, we see that the MacMahon function G$_{4}\left(
q\right) =\left\langle 0|\Psi _{+}\left( 1\right) \Omega _{-}\left( q\right)
|0\right\rangle $ eq(\ref{77}) appears as a very particular correlation
function and then cannot be the generating functional of all possible 4d
partitions.

\begin{acknowledgement}
\qquad {\small \ \ \ }\newline
{\small This research work is supported by the program Protars III D12/25.
HJ would like to thank ICTP for kind hospitality where part of this work has
been done. The authors thank B. Szendroi\ for helpful suggestion.}
\end{acknowledgement}

\section{Appendices}

\qquad In this section, we give two appendices: an \emph{Appendix A} where
we describe the vertex operators $\Gamma _{\pm }^{\left( n\right) }\left(
x\right) $ and their commutation relations algebra. An \emph{Appendix B}
which deals with the derivation of eq(\ref{cr}).

\subsection{Appendix A: Vertex operators $\Gamma _{\pm }^{\left( n\right)
}\left( x\right) $}

\qquad We first study the \emph{level n} vertex operators $\Gamma _{\pm
}^{\left( n\right) }\left( x\right) $ and their main properties starting by $%
\Gamma _{\pm }^{\left( 2\right) }=\Psi _{\pm }$. Then, we give their algebra.

\subsubsection{Level 2 vertex operator}

\qquad To begin notice that the operators $\Psi _{\pm }\left( 1\right) $ eq(%
\ref{psi-}), denoted also as\textrm{\ }$\Gamma _{\pm }^{\left( 2\right) }(1)$%
, can be put in the form, 
\begin{eqnarray}
\Psi _{-}\left( 1\right) &=&\lim_{s\rightarrow \infty }\left[ \left(
\dprod\limits_{t=0}^{s}\Gamma _{-}\left( q^{t}\right) \right) q^{sL_{0}}%
\right] ,  \notag \\
\Psi _{+}\left( 1\right) &=&\lim_{s\rightarrow \infty }\left[
q^{sL_{0}}\left( \dprod\limits_{t=0}^{s}\Gamma _{+}\left( q^{-t}\right)
\right) \right] .
\end{eqnarray}%
Using the expression of $\Gamma _{\pm }\left( z\right) $ eq(\ref{ga}), we
can rewrite $\Psi _{\pm }\left( 1\right) $ as follows: 
\begin{eqnarray}
\Psi _{-}\left( y\right) &=&\left( \dprod\limits_{t=-\infty }^{-1}\Gamma
_{-}\left( y\right) q^{L_{0}}\right) =\dprod\limits_{k=0}^{\infty }\Gamma
_{-}\left( q^{k}y\right) ,  \notag \\
\Psi _{+}\left( x\right) &=&\left( \dprod\limits_{t=0}^{\infty
}q^{L_{0}}\Gamma _{+}\left( x\right) \right) =\dprod\limits_{t=0}^{\infty
}\Gamma _{+}\left( q^{-k}x\right) ,  \label{b2}
\end{eqnarray}%
or equivalently like 
\begin{eqnarray}
\Psi _{-}\left( y\right) &=&\exp \left( \sum_{n\geq 1}\frac{i}{n}\frac{y^{n}%
}{\left( 1-q^{n}\right) }J_{-n}\right) ,  \notag \\
\Psi _{+}\left( x\right) &=&\exp \left( -\sum_{n\geq 1}\frac{i}{n}\frac{%
x^{-n}}{\left( 1-q^{n}\right) }J_{n}\right) .  \label{psi}
\end{eqnarray}%
Let us compute the algebra of these vertex operators. \newline
First, we have, 
\begin{equation}
q^{L_{0}}\Psi _{\pm }\left( z\right) q^{-L_{0}}=\Psi _{\pm }\left( qz\right)
,
\end{equation}%
showing that $q^{L_{0}}$ acts as a translation operator. We also have 
\begin{equation}
\Psi _{\pm }\left( x\right) \Psi _{\pm }\left( y\right) =\Psi _{\pm }\left(
y\right) \Psi _{\pm }\left( x\right) .
\end{equation}%
To get the commutator between $\Psi _{+}\left( x\right) $ and $\Psi
_{-}\left( y\right) $, we can do it in two ways which, by their comparison,
allow us to get a new identity:\newline
\textbf{(i)} Computation by using products of $\Gamma _{\pm }$. We have,%
\begin{eqnarray}
\Psi _{+}\left( x\right) \Psi _{-}\left( y\right) &=&\dprod\limits_{l\geq
0}\Gamma _{+}\left( q^{l}x\right) \dprod\limits_{k\geq 0}\Gamma _{-}\left(
q^{k}y\right)  \notag \\
&=&\dprod\limits_{s=0}^{\infty }\left( 1-q^{s}\frac{y}{x}\right) ^{-\left(
s+1\right) }\Psi _{-}\left( y\right) \Psi _{+}\left( x\right) ,  \label{xy}
\end{eqnarray}%
in particular 
\begin{equation}
\Psi _{+}\left( 1\right) \Psi _{-}\left( q\right) =\left(
\dprod\limits_{t=1}^{\infty }\left( 1-q^{t}\right) ^{-t}\right) \Psi
_{-}\left( q\right) \Psi _{+}\left( 1\right) .
\end{equation}%
Notice 
\begin{eqnarray}
\Gamma _{+}\left( 1\right) \Psi _{-}\left( y\right) &=&\Gamma _{+}\left(
1\right) \dprod\limits_{k\geq 0}\Gamma _{-}\left( q^{k}y\right)  \notag \\
&=&\dprod\limits_{k\geq 0}\left( 1-q^{k}y\right) ^{-1}\Psi _{-}\left(
y\right) \Gamma _{+}\left( 1\right)
\end{eqnarray}%
and also 
\begin{eqnarray}
\Gamma _{+}\left( x\right) \Psi _{-}\left( y\right) &=&\Gamma _{+}\left(
x\right) \dprod\limits_{k\geq 0}\Gamma _{-}\left( q^{k}y\right)  \notag \\
&=&\dprod\limits_{k\geq 0}\left( 1-q^{k}\frac{y}{x}\right) ^{-1}\Psi
_{-}\left( y\right) \Gamma _{+}\left( x\right) .
\end{eqnarray}%
We also have%
\begin{eqnarray}
\Psi _{+}\left( x\right) \Gamma _{-}\left( 1\right) &=&\dprod\limits_{k\geq
0}\Gamma _{+}\left( q^{-k}x\right) \Gamma _{-}\left( 1\right)  \notag \\
&=&\dprod\limits_{k\geq 0}\left( 1-q^{k}x^{-1}\right) ^{-1}\Gamma _{-}\left(
1\right) \Psi _{+}\left( x\right)
\end{eqnarray}%
\textbf{(ii)} Computation using directly eqs(\ref{psi}). We get,%
\begin{equation}
\Psi _{+}\left( x\right) \Psi _{-}\left( y\right) =\exp \left( \sum_{n\geq 1}%
\frac{1}{n}\frac{y^{n}x^{-n}}{\left( 1-q^{n}\right) ^{2}}\right) \Psi
_{-}\left( y\right) \Psi _{+}\left( x\right) .  \label{ii}
\end{equation}%
By comparing the two expressions (\ref{xy}) and (\ref{ii}), we get the
following identity 
\begin{equation}
\exp \left( \sum_{n\geq 1}\frac{1}{n}\frac{y^{n}x^{-n}}{\left(
1-q^{n}\right) ^{2}}\right) =\dprod\limits_{s=0}^{\infty }\left( 1-q^{s}%
\frac{y}{x}\right) ^{-\left( s+1\right) }.
\end{equation}%
or equivalently like,%
\begin{equation*}
\sum_{n\geq 1}\frac{1}{n}\frac{y^{n}x^{-n}}{\left( 1-q^{n}\right) ^{2}}%
=-\sum_{s=0}^{\infty }\left[ \left( s+1\right) \ln \left( 1-q^{s}\frac{y}{x}%
\right) \right]
\end{equation*}

\subsubsection{Generic q-deformed operators}

Here \ we give the expressions of the generic q-deformed operators and some
useful properties of their algebra. \newline
The starting point is the vertex operators%
\begin{equation}
\Gamma _{-}\left( z\right) =\exp \left( i\sum_{n>0}\frac{1}{n}%
z^{n}J_{-n}\right) ,\qquad \Gamma _{+}\left( z\right) =\exp \left(
-i\sum_{n>0}\frac{1}{n}z^{-n}J_{n}\right) .  \label{b1}
\end{equation}%
and the aim is: \newline
\textbf{(1)} compute for $n\geq 1$, the following hierarchy of composite
vertex operators 
\begin{equation}
\Gamma _{-}^{\left( n+1\right) }\left( z\right) =\left(
\dprod\limits_{t_{n}=1}^{\infty }\cdots \left[ \dprod\limits_{t_{2}=1}^{%
\infty }\left( \dprod\limits_{t_{1}=1}^{\infty }\Gamma _{-}\left( z\right)
q^{L_{0}}\right) q^{L_{0}}\right] \cdots q^{L_{0}}\right) .
\end{equation}%
For $n=0,$ we have just $\Gamma _{-}^{\left( 1\right) }\left( z\right)
=\Gamma _{-}\left( z\right) $. Similar quantities can be written down for $%
\Gamma _{+}^{\left( n+1\right) }\left( z\right) $; we shall not report them
here.\newline
\textbf{(2)} derive the identity (\ref{gm}).\newline
For these purposes, we proceed by using \textrm{inductive} method:

\paragraph{q- deformed vertex operators $\Gamma _{-}^{\left( 2\right)
}\left( z\right) $ and $\Gamma _{-}^{\left( 3\right) }\left( z\right) :$%
\newline
}

\textbf{(a) Case }$\Gamma _{-}^{\left( 2\right) }\left( z\right) $: \newline
In this case, we have%
\begin{equation}
\Gamma _{-}^{\left( 2\right) }\left( z\right) =\dprod\limits_{t=1}^{\infty
}\left( \Gamma _{-}\left( z\right) q^{L_{0}}\right) .
\end{equation}%
Notice that 
\begin{equation}
\Gamma _{-}^{\left( 2\right) }\left( z\right) =\lim_{s\rightarrow \infty
}\dprod\limits_{t=1}^{s}\left( \Gamma _{-}\left( z\right) q^{L_{0}}\right) .
\end{equation}%
By using the fact that $q^{L_{0}}$ acts as translation operator on $\Gamma
_{-}\left( z\right) $, we get%
\begin{equation}
\Gamma _{-}^{\left( 2\right) }\left( z\right) =\lim_{s\rightarrow \infty
}\left( \left[ \dprod\limits_{k=0}^{s}\left( \Gamma _{-}\left( q^{k}z\right)
\right) \right] q^{sL_{0}}\right) .
\end{equation}%
For simplicity, we consider the action of $\Gamma _{-}^{\left( 2\right)
}\left( z\right) $, on the vacuum, that reads as%
\begin{equation}
\Gamma _{-}^{\left( 2\right) }\left( z\right) \left\vert 0\right\rangle
=\left( \dprod\limits_{k=0}^{\infty }\Gamma _{-}\left( q^{k}z\right) \right)
\left\vert 0\right\rangle .
\end{equation}%
Substituting $\Gamma _{-}\left( q^{k}z\right) $ by its expression (\ref{b1}%
), we find%
\begin{eqnarray}
\Gamma _{-}^{\left( 2\right) }\left( z\right) \left\vert 0\right\rangle
&=&\exp \left( i\sum_{n>0}\sum_{k=0}^{\infty }\frac{q^{kn}}{n}%
z^{n}J_{-n}\right) \left\vert 0\right\rangle  \notag \\
&=&\exp \left( i\sum_{n>0}\frac{1}{n}\frac{z^{n}}{\left( 1-q^{n}\right) }%
J_{-n}\right) \left\vert 0\right\rangle .
\end{eqnarray}%
Notice that $\Gamma _{-}^{\left( 2\right) }\left( z\right) \left\vert
0\right\rangle $ can be decomposed as follows,%
\begin{equation}
\Gamma _{-}^{\left( 2\right) }\left( z\right) \left\vert 0\right\rangle
=\exp \left( i\sum_{n>0}\frac{1}{n}z^{n}J_{-n}\right) \exp \left(
i\sum_{n>0}q^{n}\sum_{k=0}^{\infty }\frac{q^{kn}}{n}z^{n}J_{-n}\right)
\left\vert 0\right\rangle
\end{equation}%
or equivalently like%
\begin{equation}
\Gamma _{-}^{\left( 2\right) }\left( z\right) \left\vert 0\right\rangle
=\exp \left( i\sum_{n>0}\frac{1}{n}z^{n}J_{-n}\right) \exp \left( i\sum_{n>0}%
\frac{1}{n}\frac{\left( qz\right) ^{n}}{\left( 1-q^{n}\right) }J_{-n}\right)
\left\vert 0\right\rangle
\end{equation}%
showing that we have:%
\begin{equation}
\Gamma _{-}^{\left( 2\right) }\left( z\right) \left\vert 0\right\rangle
=\Gamma _{-}^{\left( 1\right) }\left( z\right) \Gamma _{-}^{\left( 2\right)
}\left( qz\right) \left\vert 0\right\rangle ,\qquad z\in C.
\end{equation}%
\textbf{(b) Case }$\Gamma _{-}^{\left( 3\right) }\left( z\right) $: \newline
Here, we have%
\begin{equation}
\Gamma _{-}^{\left( 3\right) }\left( z\right) \left\vert 0\right\rangle
=\dprod\limits_{t_{2}=1}^{\infty }\left[ \Gamma _{-}^{\left( 2\right)
}\left( z\right) q^{L_{0}}\right] \left\vert 0\right\rangle
=\dprod\limits_{k=0}^{\infty }\left( \Gamma _{-}^{\left( 2\right) }\left(
q^{k}z\right) \right) \left\vert 0\right\rangle .
\end{equation}%
Substituting $\Gamma _{-}^{\left( 2\right) }\left( q^{k}z\right) $ by its
expression (\ref{b2}), we find%
\begin{eqnarray}
\Gamma _{-}^{\left( 3\right) }\left( z\right) \left\vert 0\right\rangle
&=&\exp \left( i\sum_{n>0}\frac{1}{n}\sum_{k=0}^{\infty }q^{kn}\frac{z^{n}}{%
\left( 1-q^{n}\right) }J_{-n}\right) \left\vert 0\right\rangle  \notag \\
&=&\exp \left( i\sum_{n>0}\frac{1}{n}\frac{z^{n}}{\left( 1-q^{n}\right) ^{2}}%
J_{-n}\right) \left\vert 0\right\rangle
\end{eqnarray}%
Finally, if we rewrite the above relation as follows%
\begin{equation}
\Gamma _{-}^{\left( 3\right) }\left( z\right) \left\vert 0\right\rangle
=\exp \left( \sum_{n>0}\frac{iz^{n}J_{-n}}{n\left( 1-q^{n}\right) }\right)
\exp \left( \sum_{n>0}\frac{i\left( qz\right) ^{n}J_{-n}}{n\left(
1-q^{n}\right) ^{2}}\right) \left\vert 0\right\rangle
\end{equation}%
we obtain the relation%
\begin{equation}
\Gamma _{-}^{\left( 3\right) }\left( z\right) \left\vert 0\right\rangle
=\Gamma _{-}^{\left( 2\right) }\left( z\right) \Gamma _{-}^{\left( 3\right)
}\left( qz\right) \left\vert 0\right\rangle  \label{bsy}
\end{equation}%
as claimed in section 6\textrm{\ }eq(\ref{h3}).

\paragraph{Higher levels\newline
}

\qquad From the above analysis, it is not difficult to check that the
explicit expression of the vertex operators $\Gamma _{-}^{\left( p\right)
}\left( z\right) $ acting on the vacuum is given by,%
\begin{equation}
\Gamma _{-}^{\left( p\right) }\left( z\right) \left\vert 0\right\rangle
=\exp \left( \sum_{n=1}^{\infty }\frac{iz^{n}J_{-n}}{n\left( 1-q^{n}\right)
^{p-1}}\right) \left\vert 0\right\rangle ,\quad p\geq 1.  \label{as}
\end{equation}%
Moreover using the identity,%
\begin{equation}
\frac{z^{n}}{\left( 1-q^{n}\right) ^{p-1}}=\sum_{k=1}^{p-1}\frac{\left(
qz\right) ^{n}}{\left( 1-q^{n}\right) ^{k-1}},  \label{it}
\end{equation}%
we can decompose the above relation as follows%
\begin{equation}
\Gamma _{-}^{\left( p\right) }\left( z\right) \left\vert 0\right\rangle
=\Gamma _{-}^{\left( 1\right) }\left( z\right)
\dprod\limits_{k=2}^{p-1}\Gamma _{-}^{\left( k\right) }\left( qz\right)
\left\vert 0\right\rangle .  \label{ti}
\end{equation}%
The next step is to use the relation%
\begin{equation}
\frac{z^{n}}{\left( 1-q^{n}\right) ^{p-1}}=\frac{z^{n}}{\left(
1-q^{n}\right) ^{p-2}}+\frac{\left( qz\right) ^{n}}{\left( 1-q^{n}\right)
^{p-1}},  \label{kn}
\end{equation}%
that imply the equality%
\begin{equation}
\Gamma _{-}^{\left( p\right) }\left( z\right) \left\vert 0\right\rangle
=\Gamma _{-}^{\left( p-1\right) }\left( z\right) \Gamma _{-}^{\left(
p\right) }\left( qz\right) \left\vert 0\right\rangle .
\end{equation}%
Doing the same for the first term for the right hand side%
\begin{equation}
\frac{z^{n}}{\left( 1-q^{n}\right) ^{p-2}}=\frac{z^{n}}{\left(
1-q^{n}\right) ^{p-3}}+\frac{\left( qz\right) ^{n}}{\left( 1-q^{n}\right)
^{p-2}},
\end{equation}%
the eq(\ref{kn}) can be brought to the form%
\begin{equation}
\frac{z^{n}}{\left( 1-q^{n}\right) ^{p-1}}=\frac{z^{n}}{\left(
1-q^{n}\right) ^{p-3}}+\frac{\left( qz\right) ^{n}}{\left( 1-q^{n}\right)
^{p-2}}+\frac{\left( qz\right) ^{n}}{\left( 1-q^{n}\right) ^{p-1}},
\end{equation}%
leading then to 
\begin{equation}
\Gamma _{-}^{\left( p\right) }\left( z\right) \left\vert 0\right\rangle
=\Gamma _{-}^{\left( p-2\right) }\left( z\right) \Gamma _{-}^{\left(
p-1\right) }\left( qz\right) \Gamma _{-}^{\left( p\right) }\left( qz\right)
\left\vert 0\right\rangle .
\end{equation}%
We can repeat this operation successively to end with the two following:%
\newline
(\textbf{i}) the expression of level p vertex operator as we have used it in
section 6\textrm{\ }eq(\ref{g1}), 
\begin{equation}
Z_{pd}=\left\langle 0|\mathcal{T}|0\right\rangle ,
\end{equation}%
with%
\begin{equation}
\mathcal{T}=\Gamma _{+}^{(1)}(z)\prod_{i=1}^{p_{j}}\Gamma
_{-}^{(j+1)}(q^{j}z),
\end{equation}%
where $p_{j}$ are given by 
\begin{equation}
p_{j}=\frac{\left( p-1\right) !}{j!\left( p-j-1\right) !},\qquad j=0,...,p-1.
\label{at}
\end{equation}%
(\textbf{ii}) Using the decomposition for $\Gamma _{-}^{(p)}(z)$,%
\begin{equation}
\Gamma _{-}^{(p)}(z)=\mathcal{O}_{0}\left( x_{0}\right) \mathcal{O}%
_{1}\left( x_{1}\right) \mathcal{O}_{2}\left( x_{2}\right) \cdots \mathcal{O}%
_{p-1}\left( x_{p-1}\right)
\end{equation}%
we get 
\begin{equation}
\mathcal{O}_{j+1}\left( x_{j+1}\right) =\prod_{i=1}^{p_{j}}\Gamma
_{-}^{(j+1)}(q^{j}z),\qquad j=0,...,p-1.
\end{equation}

\subsection{Appendix B: Combinatorial eq(\protect\ref{cr})}

Here we want to derive the identity (\ref{cr}) namely,%
\begin{equation}
\sum_{k=1}^{s}C_{k+p-3}^{p-2}=C_{s+p-2}^{p-1},\qquad p\geq 2.  \label{ide}
\end{equation}%
This is a standard combinatorial identity; its proof follows from basic
property \textrm{\cite{comm}},%
\begin{equation}
C_{n+1}^{k}=C_{n}^{k-1}+C_{n}^{k}\text{ }.
\end{equation}%
Applying this identity to $C_{n}^{k}$ and putting it back into the above
relation, we get,%
\begin{equation}
\begin{array}{cccc}
C_{n+1}^{k} & = & C_{n}^{k-1}+C_{n-1}^{k-1}+C_{n-1}^{k} & .%
\end{array}%
\end{equation}%
By induction, it results,%
\begin{equation}
\begin{array}{cccc}
C_{n+1}^{k} & = & \dsum\limits_{j=k-1}^{n}C_{j}^{k-1} & .%
\end{array}%
\end{equation}%
Setting $k=p-1$ \textrm{and} $n=s+p-3$, we recover the identity (\ref{ide}).%
\newline

\end{document}